\documentclass[twocolumn,aps,prb]{revtex4-2}

\usepackage[dvipdfmx]{graphicx}
\usepackage{amsmath, amssymb}
\usepackage{bm}
\usepackage{ulem}
\usepackage{mathrsfs}
\usepackage{mleftright}
\usepackage{braket}
\usepackage{color}
\usepackage{multirow}
\usepackage{physics}
\usepackage[colorlinks=true, citecolor=blue, linkcolor=blue]{hyperref}

\graphicspath{{./figures/}}

\newcommand{\diff}{\mathrm{d}}
\newcommand{\imag}{\mathrm{Im}\,}
\newcommand{\imu}{\mathrm{i}}
\newcommand{\epn}{\mathrm{e}}
\newcommand{\sgn}{\mathrm{sgn}\,}
\newcommand{\ua}{\uparrow}
\newcommand{\da}{\downarrow}
\newcommand{\dg}{\dagger}
\newcommand{\la}{\langle}
\newcommand{\ra}{\rangle}
\newcommand{\al}{\alpha}
\newcommand{\sg}{\sigma}
\newcommand{\gm}{\gamma}
\newcommand{\ep}{\varepsilon}

\begin{document}

\title{Symmetry-breaking perturbations in the 
Jahn-Teller-Hubbard model
}

\author{
Natsuki Okada$^{1,2}$, Tatsuya Miki$^{2,1 }$,
Yusuke Nomura$^{3,4}$, 
Philipp Werner$^{5}$ 
and Shintaro Hoshino$^{1,2}$
}

\affiliation{
$^1$Department of Physics, Chiba University, Chiba 263-8522, Japan
\\
$^2$Department of Physics, Saitama University, Saitama 338-8570, Japan
\\
$^3$Institute for Materials Research (IMR), Tohoku University, Sendai, 980-8577, Japan
\\
$^5$Advanced Institute for Materials Research (WPI-AIMR), Tohoku University, Sendai, 980-8577 Japan
\\
$^5$Department of Physics, University of Fribourg, 1700 Fribourg, Switzerland
}

\date{\today}

\begin{abstract}
We study the effect of symmetry-breaking perturbations in the multiorbital Hubbard model coupled to anisotropic Jahn-Teller phonons, which is relevant for the description of fulleride superconductors.
This system is often approximated by a model with static antiferromagnetic (AFM) Hund's coupling, in which the coupling to the Jahn-Teller phonon is effectively described, but 
the retardation effect associated with phonon propagation is neglected.
We compare the properties of the models with static AFM Hund's coupling and dynamical Jahn-Teller electron-phonon interaction by means of the Eliashberg theory.
Considering the susceptibilities for the spin, magnetic orbital, electric orbital, and superconductivity, we reveal a qualitatively different behavior between the two models in the case of the magnetic orbital susceptibility.
We further study the effect of a magnetic field on the $s$-wave spin-singlet superconducting state.
In the presence of the field, the magnetic orbital susceptibility becomes nonzero due to a combination of multiorbital and retardation effects, while the spin susceptibility remains zero at low temperatures.
By analyzing this phenomenon both numerically and analytically, we clarify that odd-frequency
pairs induced by the magnetic field play a crucial role in the spin and orbital magnetic susceptibilities.
Thus, the magnetic 
degrees of freedom produce interesting behaviors in the presence of retardation effects associated with 
electron-phonon coupling.
\end{abstract}

\maketitle

\section{Introduction}

Alkali-doped fullerides are strongly correlated electron systems with active molecular $t_{1u}$ orbital degrees of freedom. They exhibit a high transition temperature of up to 38 K to an $s$-wave superconducting state \cite{Gunnarsson97,Capone09,Nomura16,Takabayashi16}. The anisotropic molecular vibrations, known as Jahn-Teller phonons, play an essential role in driving the unique phenomena observed in fullerides. Specifically, the coupling between the electrons and the Jahn-Teller phonons in these multiorbital systems induces an effectively antiferromagnetic (AFM) Hund's coupling \cite{Fabrizio97,Capone02,Nomura15}.

The superconducting and Mott insulating phases in electron-phonon coupled models have been extensively studied  \cite{Han00,Han03,Yamazaki13,Nomura15_2,Kaga22}.
Analyses of the multiorbital Hubbard model with effective antiferromagnetic Hund's coupling, and hence without an explicit description of the phonons, have also been performed  \cite{Capone04, Hoshino16, Steiner15, Steiner16, Hoshino17, Misawa17, Ishigaki18, Yue21,Iwazaki21}.
While the effective AFM Hund’s coupling simplifies the theoretical analysis, electron–phonon coupling leads to a retardation effect, meaning that phonon-mediated electronic interactions are delayed due to the phonon propagation.
Hence, it is necessary to investigate how accurately the effective Hund's coupling in the static limit can describe the superconductivity, as well as magnetic and orbital properties.
This paper aims to elucidate the impact of the retardation effect from the coupling to Jahn-Teller phonons on normal and superconducting states.

To approach this problem, we employ the dynamical mean-field theory (DMFT) \cite{Georges96} combined with the multiorbital Eliashberg theory.
In DMFT, the self-energy is approximated to be spatially local, which becomes exact in infinite dimensions. 
Since A$_3$C$_{60}$ forms cubic lattices in three dimensions \cite{Gunnarsson97,Capone09,Nomura16,Takabayashi16}, we expect that the DMFT should be a good approximation for fullerides.
As for the Eliashberg approach, the theoretical framework is based on the perturbation theory \cite{Kaga22}, and hence is valid in the weak coupling regime.
While the fullerides show Mott insulating behavior for a large volume of the unit cell, our approach is valid in the opposite limit of small volume, where superconductivity is realized at low temperatures.

In our previous study \cite{Kaga22} that incorporated phonon retardation effects using multiorbital Eliashberg theory, the contribution of the phonons to superconductivity has been thoroughly analyzed in a symmetric phase \cite{Kaga22}.
In the present study, we extend the multiorbital Eliashberg equations to a framework that allows the calculation 
in low-symmetry situations, evaluating spin, orbital, and superconducting susceptibilities. 
Comparing the susceptibilities
between the two cases with and without retardation of the phonon-mediated interactions in the normal state,
the susceptibilities of the orbital electric quadrupole and $s$-wave superconductivity show qualitatively similar behaviors. 
On the other hand, we find that qualitative differences are observed in the orbital magnetic susceptibility, which indicates the importance of the retardation effect for the magnetic orbital degrees of freedom.

We also study the magnetic susceptibilities in the $s$-wave spin-singlet superconducting state.
The contribution to the susceptibility is separated into spin ($\bm{\mathcal S}$) and magnetic orbital ($\bm {\mathcal L}$) components. The spin susceptibility is zero at low temperatures, while a nonzero magnetic orbital susceptibility appears. 
We elucidate, using both numerical and analytic approaches, that the multiorbital nature and retardation effect are essential for the nonzero magnetic orbital susceptibility, and that odd-frequency pairing \cite{Berezinskii74,Kirkpatrick91,Emery92,Balatsky92,Coleman93,Tanaka12,Linder19} induced by the spin Zeeman effect is required for the vanishing spin response.
While preliminary results have appeared in our previous work~\cite{Okada_proc}, the present paper provides a detailed analysis and clarifies the physical origin of such anomalous behaviors of the magnetic susceptibilities.

We organize the paper as follows. In the next section, we present the theoretical framework of the DMFT+Eliashberg approach. Section~\ref{sec:results} is devoted to the numerical results, and Sec.~\ref{sec:susceptibilities} provides a detailed study of the magnetic susceptibilities. 
A summary is given in Sec.~\ref{sec:summary}. 
The appendices include useful formula for analyzing the multiorbital Eliashberg equations.

\section{Model and Formalism}

\subsection{Jahn-Teller-Hubbard model \label{sec:model}}

We consider a multiorbital electron model coupled to isotropic and anisotropic molecular vibrations of the fullerene \cite{Kaga22}. The total Hamiltonian is given by $\mathscr{H} = \mathscr{H}_{0e} + \mathscr{H}_{0p} + \mathscr{H}_{\mathrm{C}} + \mathscr{H}_{ep}$, where the non-interacting part $\mathscr{H}_{0}=\mathscr{H}_{0e}+\mathscr H_{0p}$ reads
\begin{align}
    \mathscr{H}_{0e} 
    &= 
    \sum_{ij}\sum_{\gm\gm'\sg} 
    \left(  t_{ij}^{\gm\gm'} -\mu \delta_{ij} \delta_{\gm\gm'} 
    \right) c_{i\gm\sg}^\dg c_{j\gm'\sg},
    \nonumber \\
    \mathscr H_{0p} &=
    \sum_{i}\sum_{\substack{\eta=0,1,3,\\4,6,8}}\omega_\eta a_{i\eta}^\dg a_{i\eta}.
    \label{eq:non_int}
\end{align}
The first term $\mathscr H_{0e}$ represents the kinetic energy of the electrons, where $i, j$ are site indices, $\sigma = \ua, \da$ denotes the spin index, and $\gamma = x, y, z$ labels the $t_{1u}$ molecular orbitals, which resemble atomic $p$ orbitals. 
The chemical potential is denoted by $\mu$.
The second term $\mathscr H_{0p}$ corresponds to the 
energy of local phonons. Here, $\eta$ labels the vibrational modes of the fullerene molecule: the isotropic vibrational mode (the irreducible representation $A_g$ of the molecular point group $I_h$) is denoted by $\eta = 0$ and the anisotropic vibrational modes ($H_g$) correspond to $\eta = 1, 3, 4, 6, 8$.
While there are several vibration modes coupled to the $t_{1u}$ electron orbitals, here we consider the representative ones with the largest electron-phonon coupling \cite{Kaga22}.

The Coulomb interaction term $\mathscr{H}_{\mathrm{C}}$ is given by
\begin{align}
  \mathscr{H}_{\mathrm{C}}
  &=
    \frac{1}{2} 
    \sum_{i\sg\sg'}\sum_{\gm_1\gm_2\gm_3\gm_4}
    U_{\gm_1\gm_2\gm_3\gm_4}
    c_{i\gm_1\sg}^\dg c_{i\gm_2\sg'}^\dg c_{i\gm_4\sg'} c_{i\gm_3\sg} 
    \nonumber\\
  &=
  U\sum_{i\gm}n_{i\gm\ua} n_{i\gm\da}
  +
  \frac{U'}{2} \sum_{i\sg\sg'}\sum_{\gm\neq\gm'} n_{i\gm\sg}n_{i\gm'\sg'}
  \nonumber\\
  &\hspace{5mm}
  +
  \frac{J}{2} \sum_{i\sg\sg'}\sum_{\gm\neq\gm'} c_{i\gm\sg}^\dg c_{i\gm'\sg'}^\dg c_{i\gm\sg'} c_{i\gm'\sg}
  \nonumber\\
  &\hspace{10mm}
  +
  \frac{J'}{2} \sum_{i\sg}\sum_{\gm\neq\gm'} c_{i\gm\sg}^\dg c_{i\gm\bar\sg}^\dg c_{i\gm'\bar\sg} c_{i\gm'\sg},
  \label{eq:Coulomb_int_SKform}
\end{align}
where the interaction is parametrized as the Slater-Kanamori interaction: 
$U_{\gamma\gamma\gamma\gamma} = U$ represents the intra-orbital interaction, $U_{\gamma\gamma'\gamma'\gamma} = U'$ denotes the inter-orbital interaction, and $U_{\gamma\gamma'\gamma\gamma'} = U_{\gamma\gamma\gamma'\gamma'} = J$ ($\gamma \neq \gamma'$) corresponds to the ferromagnetic Hund's coupling. Assuming spherical symmetry for the interactions, we set $U' = U - 2J$.
The number operator is defined by $n_{i\gm\sg} = c_{i\gm\sg}^\dg c_{i\gm\sg}$.

\begin{figure}[tb]
    \centering
    \includegraphics[width=85mm]{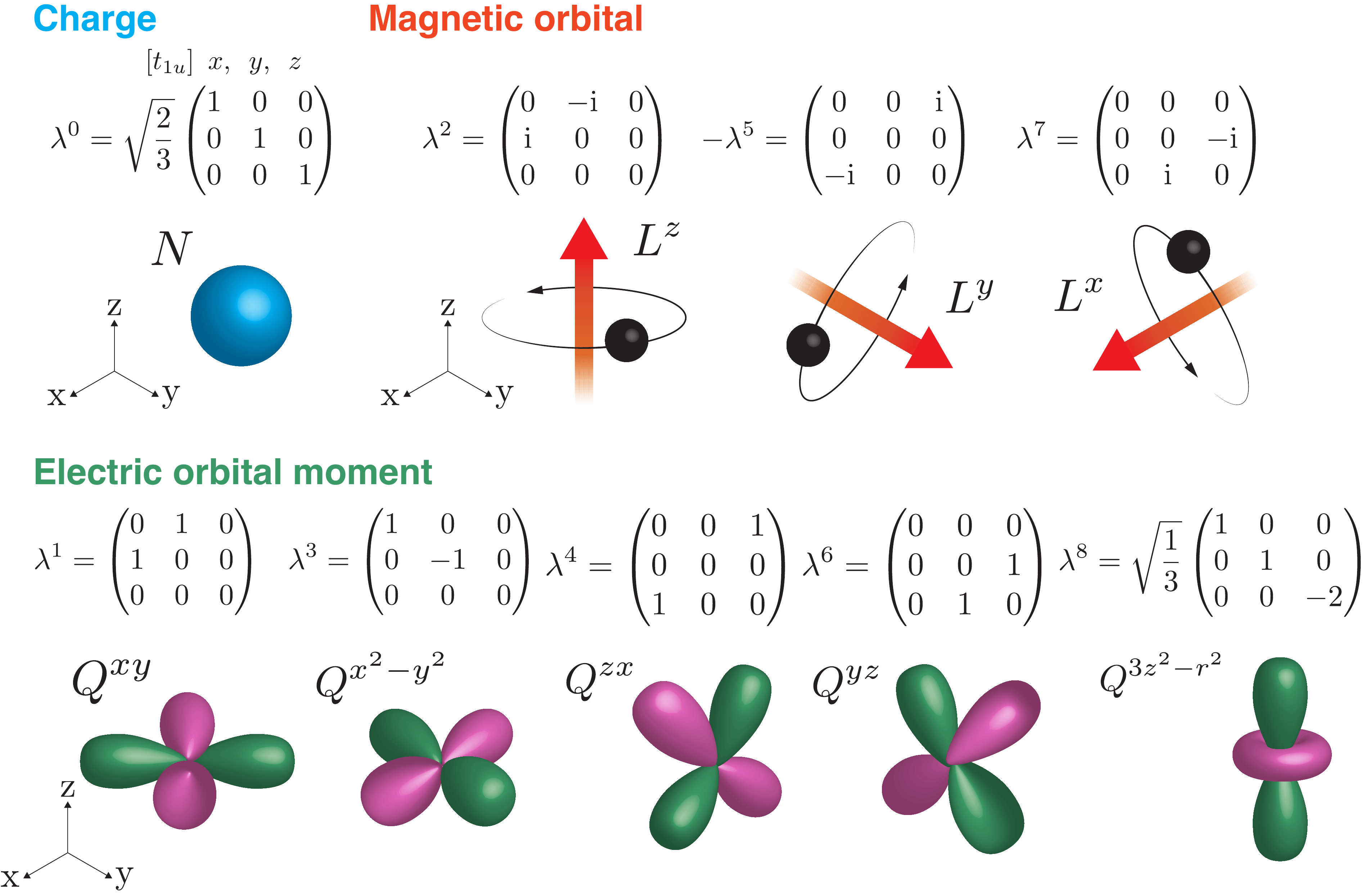}  
    \caption{
    Definition of the Gell-Mann matrices and their schematic illustration in terms of electric (monopole $N$ and quadrupole $Q$) and magnetic (dipole $L$) physical quantities. 
   }
    \label{fig:Gell-Mann}
\end{figure}

By introducing the charge-orbital moment $T_{i\eta}$ [see Eq.~\eqref{eq:charge-orbital_moment}],
the interaction in Eq.\,\eqref{eq:Coulomb_int_SKform} can be expressed in a simplified form \cite{Iimura21}: 
\begin{align}
  \mathscr{H}_{\mathrm{C}}
  &=
  \sum_{i} \sum_{\substack{\eta=0,1,3,\\4,6,8}}
  I_\eta : T_{i\eta}T_{i\eta} :, \label{eq:ham_int_I}
\end{align}
where the colon symbol ( $:$ ) represents the normal ordering, and $I_0=3U/4-J$, $I_{1,3,4,6,8}=J/2$. 
The charge-orbital moment $T_{i\eta}$ is defined as
\begin{align}
  T_{i\eta}
  &=
  \sum_{\gm\gm'\sg}
  c_{i\gm\sg}^\dg \lambda_{\gm\gm'}^\eta c_{i\gm'\sg}, 
  \label{eq:charge-orbital_moment} 
\end{align}
where $\lambda^\eta$ are the Gell-Mann matrices
displayed in Fig.~\ref{fig:Gell-Mann}.
Physically, $\lambda^0$ represents charge degrees of freedom, $\lambda^{2,5,7}$ the magnetic orbital moment (i.e., orbital angular momentum or magnetic dipole), and $\lambda^{1,3,4,6,8}$ the electric orbital moment (i.e., electric quadrupole).
Their symmetry is represented by the polynomials.
For the $\eta=0$ component, $r^2=x^2+y^2+z^2$ is assigned, corresponding to the irreducible representation $A_1$ in the cubic point group 
of A$_3$C$_{60}$.
The anisotropic orbital or quadrupolar moments are characterized by $\eta= 1$ ($xy$ type), $\eta=6$ ($yz$ type), $\eta = 4$ ($zx$ type), $\eta=8$ ($3z^2-r^2$ type), and $\eta=3$ ($x^2-y^2$ type), which belong to the irreducible representations 
$T_2$ and $E$.
Note that the electronic ($\eta=0,1,3,4,6,8$) and magnetic ($\eta=2,5,7$)  degrees of freedom can also be understood from the irreducible representations for the symmetric product $[t_{1u}\times t_{1u}] = A_{1} + T_{2} + E$ and anti-symmetric product $\{t_{1u}\times t_{1u}\} = T_1$, respectively.

The electron-phonon interaction part $\mathscr{H}_{ep}$ is given by
\begin{align}
  \mathscr{H}_{ep}
  &=  \sum_{i}\sum_{\substack{\eta=0,1,3,\\4,6,8}}  
  g_\eta \phi_{i\eta} T_{i\eta}, 
  \label{eq:e-ph_int}
\end{align}
where $\phi_{i\eta} = a_{i\eta} + a_{i\eta}^\dg$ is a displacement operator of the fullerene molecules, and $g_\eta$ is the electron-phonon coupling constant. 
We note the absence of $\eta = 2$ ($z$ type),  $\eta = 5$ ($y$ type), and  $\eta = 7$ ($x$ type) components in the Hamiltonian, due to their magnetic nature, which prohibits a coupling to the molecular structure displacement.
On the other hand, these antisymmetric $\eta=2,5,7$ 
components will be used for the external field part (see Sec.~\ref{sec:External_perturbations}).

\subsection{Eliashberg equation in spin, orbital, and Nambu space}

In this section, we construct a multiorbital Eliashberg equation based on the Jahn-Teller-Hubbard model.
Since the fulleride superconductors possess three-dimensional cubic structures, the DMFT approach is suitable for their theoretical analysis, in which the self-energy is approximated by a spatially local quantity.
The problem is mapped onto an effective impurity problem, where the fullerene molecule is embedded in a self-consistently determined dynamical environment.
We solve the impurity problem by the perturbation theory, leading to the local Eliashberg equation \cite{Kaga22}.
In order to study the various susceptibilities, we consider the solution under any kinds of external fields, for which we need to consider arbitrarily low-symmetric solutions.

This subsection is structured as follows.
The Green's functions are introduced in Secs.~\ref{sec:electron_green_function} and \ref{sec:phonon_green_function}. 
The shift of the vibrational center of the phonons is explained in Sec.~\ref{sec:shift_phonon}. 
The self-energies related to the Green's functions are discussed in Secs.~\ref{sec:self-energies_electron-phonon} and \ref{sec:self-energies_coulomb}.
A short summary of this subsection is provided in Sec.~\ref{sec:summary_Eliashberg}.

\subsubsection{Electron Green's function \label{sec:electron_green_function}}

We first introduce the electron Green's function.
By using the twelve-component Nambu spinor 
\begin{align}
\bm \Psi &= (c_{x\ua},c_{x\da},c_{y\ua},c_{y\da},c_{z\ua},c_{z\da},c^\dg_{x\ua},c^\dg_{x\da},c^\dg_{y\ua},c^\dg_{y\da},c^\dg_{z\ua},c^\dg_{z\da})^\mathrm{T}
    , \label{eq:nambu_spinor}
\end{align}
the local Green's function is defined as follows:
\begin{align}
  \check{G}(\tau) 
    &= 
    -\la \mathcal{T}\,\bm\Psi(\tau)\bm\Psi^\dg \ra
    =
    \begin{pmatrix}
        \hat{G}(\tau) & \hat{F}(\tau) \\
        \hat{\bar{F}} (\tau) & \hat{\bar{G}}(\tau)
    \end{pmatrix},  \label{eq:gcheck}
\end{align}
where the site index is omitted by assuming translational symmetry within the solid. 
We employ this omission in the rest of this paper unless otherwise stated.
The $6\times6$ matrix in spin-orbital space is denoted by the hat symbol (~$\hat{\ }~$), while the $12\times12$ matrix 
in Nambu space is denoted by the check symbol (~$\check{\ }$~). 
In Eq.~\eqref{eq:gcheck}, $A(\tau) = \epn^{\tau\mathscr{H}} A  \epn^{-\tau\mathscr{H}}$ is the Heisenberg representation of the operator in imaginary time, and $ \mathcal{T}$ denotes the imaginary-time ordering operator.
The component representation of the Green's function matrix is explicitly given by
\begin{align}
        \hat{G}_{\gm\sg,\gm'\sg'}(\tau)
        &= - \la \mathcal T c_{\gm\sg}(\tau) c_{\gm'\sg'}^\dg \ra
        \label{eq:hat_G_def},
        \\
        \hat{F}_{\gm\sg, \gm'\sg'}(\tau) 
        &= - \la \mathcal T c_{\gm\sg}(\tau) c_{\gm'\sg'} \ra
        ,
        \\
        \hat{\bar{F}}_{\gm\sg, \gm'\sg'}(\tau) &= - \la \mathcal T c^\dg_{\gm\sg}(\tau) c_{\gm'\sg'}^\dg \ra
        ,\\
        \hat{\bar{G}}_{\gm\sg, \gm'\sg'}(\tau)
        &= - \la \mathcal T c_{\gm\sg}^\dg(\tau) c_{\gm'\sg'} \ra
        .
\end{align}
The local Green's function is obtained 
once the local self-energy $\check{\Sigma}(\imu\omega_n)$ is given. 
The explicit relationship between them is given by 
\begin{align}
     &\check{G}(\imu\omega_n) 
    = 
    \int_{-D}^{D} \diff\ep 
    \rho(\ep)
    \left[
    \imu\omega_n \check{1} - (\ep - \mu)\check{\tau}_3 - \check{\Sigma}(\imu\omega_n)
    \right]^{-1}, 
    \label{eq:G_loc_def}
\end{align}
with $\check{\tau}_3 = \mathrm{diag}(\hat{1}, -\hat{1})$ and the fermionic Matsubara frequency $\omega_n = (2n+1)\pi T$ ($n\in\mathbb{Z}$).
The integral over $\ep$ corresponds to the wave-vector integral needed for obtaining the local Green's function.
For simplicity, the density of states is chosen to have a featureless semi-elliptical shape:
\begin{align}
    \rho(\ep) = 
    \frac{2}{\pi D^2}\sqrt{D^2 - \ep^2}, 
\end{align}
where the bandwidth is given by $2D$.
We note that the $\ep$-integral can be performed analytically as detailed in Appendix \ref{sec:calc_green_function_app}.

\subsubsection{Phonon Green's function \label{sec:phonon_green_function}}

The local phonon Green's function is defined in terms of the displacement operator $\phi_\eta$:
\begin{align}
    D_{\eta\eta'}(\tau) &= -\la\mathcal{T}\, \phi_\eta(\tau) \phi_{\eta'} \ra , \label{eq:green_phonon}
\end{align}
which forms a $6\times6$ matrix with respect to $\eta$ and $\eta'$.

Given the local phonon self-energy $\Pi_{\eta\eta'}(\imu\nu_m)$, the phonon Green's function in Eq.~\eqref{eq:green_phonon} can be computed by using the Dyson equation: 
\begin{align}
    D_{\eta\eta'}^{-1}(\imu\nu_m) &= D_{0, \eta\eta'}^{-1}(\imu\nu_m) - \Pi_{\eta\eta'}(\imu\nu_m),
    \label{eq:def_phonon_Green_Matsubara}
    \\
    D_{0, \eta\eta'}(\imu\nu_m)
    &=
    \frac{2\omega_\eta}{(\imu\nu_m)^2 - \omega_\eta^2}\delta_{\eta\eta'}, \label{eq:d0}
\end{align}
where $\nu_m = 2m\pi T$ ($m\in\mathbb{Z}$) denotes the bosonic Matsubara frequencies.

\subsubsection{Shift of the phonon displacement operator \label{sec:shift_phonon}}

Before deriving the Eliashberg equation, which connects the Green's functions to the self-energies, we discuss the shift of the vibration center of the phonons.
This procedure is related to the removal of the tadpole diagrams in perturbation theory.

Since we are interested in a low-symmetric system under external fields, we should be careful about nonzero 
$\la T_\eta\ra$ in general.
In the language of Feynman diagrams, it generates tadpole diagrams originating from electron-phonon coupling.
In order to see this, let us employ the path-integral formalism \cite{Kaga22}.
The phonon degrees of freedom are integrated out without loss of generality, to yield an effective electron-electron interaction term in the action:
\begin{align}
    S_1 &= \frac 1 2\sum_{\eta\eta'} g_\eta g_{\eta'} \int \diff \tau \diff \tau' D_{0,\eta\eta'} (\tau-\tau') T_\eta(\tau) T_{\eta'} (\tau') \label{eq:action_s1}
    , 
\end{align}
where $\tau$ denotes imaginary time. 
In this formalism, the charge-orbital moments corresponding to Eq.~\eqref{eq:charge-orbital_moment} are defined by
\begin{align}
    T_\eta(\tau) &= \sum_{\sg \gm \gm'}\bar c_{\gm\sg}(\tau) \lambda^\eta_{\gm\gm'} c_{\gm'\sg}(\tau)
    ,
\end{align}
where $c$ and $\bar c$ are Grassmann numbers associated with the electron's annihilation and creation operators.
In order to avoid the explicit appearance of the tadpole diagrams,
we subtract the mean value of $T_\eta$ in advance [see also Eq.~\eqref{eq:charge-orbital_ave}].
Namely, we define
\begin{align}
    \delta T_\eta(\tau) = T_\eta (\tau) - \la T_\eta \ra
\end{align}
and rewrite the action in Eq.~\eqref{eq:action_s1} as
\begin{align}
    S_1 &= \frac 1 2\sum_{\eta\eta'} g_\eta g_{\eta'} \int \diff \tau \diff \tau' D_{0,\eta\eta'}(\tau-\tau') \delta T_\eta(\tau) \delta T_{\eta'} (\tau')
    \nonumber \\
    & \hspace{5mm} + \sum_\eta F_\eta  \int \diff \tau    
    \delta T_\eta(\tau)
    , \label{eq:S1_effective}
\end{align}
where the following static orbital field appears:
\begin{align}
    F_\eta &= g_\eta \sum_{\eta'} g_{\eta'}\la T_{\eta'}\ra 
    D_{0,\eta\eta'} (\imu\nu = 0)
    \label{eq:Hartree}.
\end{align}
If the emergent potential $F_\eta$ is included in the electron 
self-energy as a static contribution, 
the tadpole diagram contribution can be dropped. 
The static charge-orbital moment is 
evaluated from the Green's function as
\begin{align}
    \la T_\eta\ra &= - \sum_{\gm\gm'\sg} \lambda^\eta_{\gm\gm'} \hat G_{\gm'\sg,\gm\sg} (\tau = -0^+) \label{eq:charge-orbital_ave}
    ,
\end{align}
where $\hat G$ is defined in Eq.~\eqref{eq:hat_G_def}.

\begin{figure}[tb]
    \centering 
    \includegraphics[width=75mm]{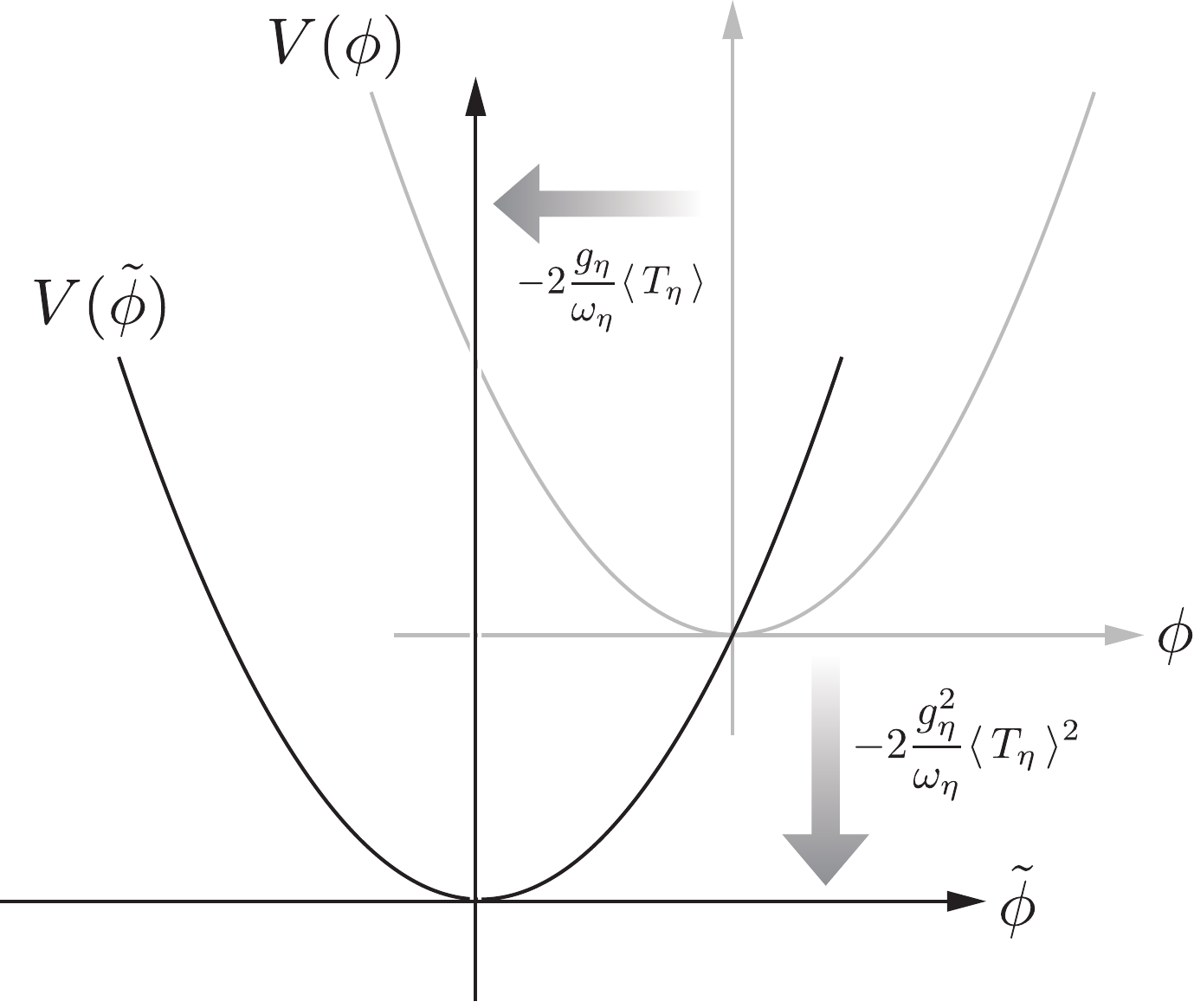}  
    \caption{
   Schematic picture of the shift of harmonic potential in the presence of the finite quadrupolar moment.
   The original potential $V(\phi_\eta )= \tfrac{\omega_\eta}{4}\phi_\eta^2$ is shifted to $V(\phi_\eta ) + g\phi_\eta \la T_\eta\ra = V(\tilde \phi_\eta ) - \frac{2g_\eta^2 \la T_\eta^2\ra}{\omega_\eta}$ in the presence of a static charge-orbital moment $\la T_\eta \ra$, where $\tilde \phi_\eta = \phi_\eta + \frac{2g_\eta \la T_\eta\ra}{\omega_\eta}$.
    }
    \label{fig:shift}
\end{figure}

The above argument in terms of the imaginary-time action formalism can be interpreted intuitively as a shift of the phonon displacement operator, which is analogous to the Lang-Firsov transformation \cite{Lang62}.
In the presence of the orbital moment $\la T_\eta \ra$, the vibration center of the displacement $\phi_{\eta}$ is shifted as schematically illustrated in Fig.~\ref{fig:shift}.
To account for this effect explicitly, we rewrite the phonon Hamiltonian as
\begin{align}
    \mathscr H_{0p} + \mathscr H_{ep}
    &= \sum_\eta \qty[ \frac{\omega_\eta}{4}(p_\eta^2 + \phi_\eta^2-2) + g_\eta \phi_\eta T_\eta  ], 
\end{align}
where $p_\eta = (a_\eta - a_\eta^\dg)/\imu$ is the momentum operator (the site index is omitted for clarity).
Note that here we employ the operator formalism.
We define the shifted operators by
\begin{align}
    \tilde \phi_\eta &= \phi_\eta + \frac{2g_\eta \la T_\eta \ra}{\omega_\eta},
    \\
    \tilde a_\eta &= (\tilde \phi_\eta + \imu p_\eta)/2,
\end{align}
which gives
\begin{align}
    \mathscr H_{0p} + \mathscr H_{ep}
    &= \sum_\eta \qty( \omega_\eta \tilde a_\eta^\dg \tilde a_\eta + g_\eta \tilde \phi_\eta \delta T_\eta 
    - \frac{2g_\eta^2}{\omega_\eta} \la T_\eta \ra \delta T_\eta
    ) .
\end{align}
The last term of this equation is equivalent to the last term in Eq.~\eqref{eq:S1_effective}.
Note that $\tilde a_\eta$ satisfies the standard commutation relation for bosons.
The zeroth-order Green's function for $\tilde \phi_\eta$ is identical to the one for $\phi_\eta$ in the absence of the shift.

The merit of the path-integral approach is that the phonon Green's function is directly involved and it is easier to account for the phonon self-energy effects in the perturbation theory.
The relation to the Green's function is not apparent in the Hamiltonian formalism.

\subsubsection{Self-energies from electron phonon coupling: Eliashberg equation \label{sec:self-energies_electron-phonon}}

We proceed with the discussion of the self-energies for the electrons and phonons, i.e. the  Eliashberg equation.
We employ the lowest-order perturbation theory in the Green's function formalism.
The second-order self-energies derived from electron-phonon coupling are given by
\begin{align}
    \check \Sigma_{\mathrm{ep}} (\tau) &=
    - 
    \sum_{\eta, \eta'}
    g_\eta g_{\eta'}D_{\eta\eta'} (\tau)
    \Bigl[
        \check{\lambda}^{\eta} 
        \check{G}^{\eta'}(\tau)
    \Bigr], 
    \label{eq:ep_Eliashberg}
    \\
    \Pi_{\eta\eta'}(\tau) &=
    \frac 1 2
    g_{\eta}g_{\eta'}  
    \Tr\,
    \Bigl[
        \check{G}^{\eta'}(- \tau)
        \check{G}^{\eta}(\tau)
    \Bigr].  \label{eq:ep_phonon}
\end{align}
In the above expression, we have introduced the short-hand notation $\check G^\eta (\tau) = \check G (\tau )\check \lambda^\eta$ with
\begin{align}
    \check{\lambda}^\eta =
    \begin{pmatrix}
        \hat{\lambda}^\eta & 0 \\
        0 & -\hat{\lambda}^{\eta {\rm T}}
    \end{pmatrix} \label{eq:exGell-Mann}
\end{align}
and $\hat{\lambda}^\eta = \lambda^\eta \otimes 1_{2\times2}$.
Diagrammatically, Eq.~\eqref{eq:ep_Eliashberg} corresponds to the Fock term. The Hartree contribution has already been accounted for in the previous subsection (Sec.~\ref{sec:shift_phonon}).

\subsubsection{Self-energies from Coulomb repulsive interaction \label{sec:self-energies_coulomb}}
For the Coulomb interaction part of the self-energy $\check \Sigma_{\mathrm{C}}$, we consider the lowest-order contribution.
Here, we derive $\check \Sigma_{\mathrm{C}}$ using the variational principle \cite{Misu23}.
To begin with, we introduce the variational Hamiltonian:
\begin{align}
    \mathscr{H}_{\mathrm{var}}
    &=
    \sum_{i\sg\sg'}\sum_{\gm\gm'}
    \hat E_{\gm\sg, \gm'\sg'} c_{\gm\sg}^\dg c_{\gm'\sg'}
    \nonumber\\
    &\hspace{5mm}
    +
    \frac{1}{2} \sum_{i\sg\sg'}\sum_{\gm\gm'}
    \left( 
    \hat W_{\gm\sg, \gm'\sg'} c_{\gm\sg}^\dg c_{\gm'\sg'}^\dg + \mathrm{H. c. }
    \right). 
    \label{eq:var_ham}
\end{align}
We determine the parameters $\hat E_{\gm\sg,\gm'\sg'}$ and $\hat W_{\gm\sg,\gm'\sg'}$ by minimizing the energy difference $\la \mathscr{H}_{\mathrm{var}} - \mathscr{H}_{\mathrm{C}} \ra_0$, where the average $\la \cdots \ra_0$ is taken with $\mathscr{H}_{\mathrm{var}}$.
Thus, the parameters $\hat E_{\gm\sg,\gm'\sg'}$ and $\hat W_{\gm\sg,\gm'\sg'}$ are obtained from the relations
\begin{align}
    \hat E_{\gm\sg,\gm'\sg'} &= 
    \frac{\partial \la \mathscr H_{\mathrm{C}} \ra_0}{\partial \la c^\dg_{\gm\sg} c_{\gm'\sg'}\ra_0}, 
    \\
    \hat W_{\gm\sg,\gm'\sg'} &= 
     \frac{\partial \la \mathscr H_{\rm C} \ra_0}{\partial \la c^\dg_{\gm\sg} c^\dg_{\gm'\sg'}\ra_0}. 
\end{align}
Then, we obtain the self-energy as
\begin{align}
    \check{\Sigma}_{\rm C}
    =&    
    \begin{pmatrix}
        \hat{E} & \hat{W} \\
        \hat{W}^\dg & -\hat E^{\rm T}
    \end{pmatrix}
    , \label{eq:sigma_coulomb} \\
    \hat E_{\gm\sg, \gm'\sg'}
    =& \,
    2I_\eta \lambda_{\gm\gm'}^\eta \delta_{\sg\sg'}
    \Tr
    \left[
        \hat{G}(\tau=-0^+) \hat{\lambda}^\eta
    \right]
    \nonumber\\
    &
    \, -
    2I_\eta 
    \left[
        \hat{\lambda}^\eta \hat{G} (\tau=-0^+)  \hat{\lambda}^\eta
    \right]_{\gm\sg, \gm'\sg'}, \label{eq:Coulomb1}
    \\
    \hat W_{\gm\sg, \gm'\sg'}
    =& \,
    2I_\eta 
    \left[
        \hat{\lambda}^\eta \hat{F}(\tau=-0^+) \hat{\lambda}^\eta
    \right]_{\gm\sg, \gm'\sg'},  \label{eq:Coulomb2}
\end{align}
where the first and second terms in Eq.~\eqref{eq:Coulomb1} are the Hartree and Fock terms, respectively.

The Coulomb repulsion on the $t_{1u}$ orbitals leads to the realization of electronic states that minimize the energy, which tends to decrease the number of electrons. 
Hence, the chemical potential should be adjusted to ensure that the electron number is $n=3$ at half filling, and it is convenient to consider the corresponding energy shift in advance.
To estimate this contribution, we choose
\begin{align}
    \la c_{\gm\sg}^\dg c_{\gm'\sg'}\ra \to \frac{n}{2\times3} \delta_{\gm\gm'}\delta_{\sg\sg'}. 
\end{align}
This provides the explicit form of the Hartree term $\hat E^{(0)}$ as
\begin{align}
    \hat E^{(0)}_{\gm\sg, \gm'\sg'}
    &= \frac{20}{3}(I_0 - I_1)\delta_{\gm\gm'} \delta_{\sg\sg'}\times\frac{n}{6}, 
\end{align}
where we have used the relations for the Gell-Mann matrices $\sum_{\eta=0}(\lambda^\eta)^2= \frac{2}{3}\delta_{\gm\gm'}$, $\sum_{\eta=8,3}(\lambda^\eta)^2 = \frac{4}{3}\delta_{\gm\gm'}$, $\sum_{\eta=1,6,4}(\lambda^\eta)^2=2\delta_{\gm\gm'}$.
By subtracting the Hartree-induced chemical potential shift from the beginning, we can perform the computation so that the electron number remains at $n = 3$ when the electron-hole symmetry is preserved.

\subsubsection{Summary of the Eliashberg equation \label{sec:summary_Eliashberg}}

In the above, we have derived the self-energies, namely the Eliashberg equation.
Since the above equations have a complicated structure, 
we briefly summarize them here.

In this paper, we consider the Green's function of the electrons $\check G(\tau)$ in the basis of the extended Nambu space (2 spins, 3 orbitals, Nambu) [see Eq.~\eqref{eq:nambu_spinor}] defined by Eq.~\eqref{eq:gcheck}, in order to investigate low-symmetric systems in the presence of external fields.
We also consider the Green's function of the phonons in the $\eta$-space ($\eta = 0,1,3,4,6,8$), denoted by $D(\tau)$ [Eq.~\eqref{eq:green_phonon}].

The self-energy of the electrons $\check \Sigma$ is derived from the electron-phonon interaction and Coulomb interaction, which takes into account the lowest order of the perturbation.
The concrete expressions are given by Eqs.~\eqref{eq:ep_Eliashberg} and \eqref{eq:sigma_coulomb}-\eqref{eq:Coulomb2}.
We also consider the self-energy of the phonons $\Pi$ within second-order perturbation theory, as written in Eq.~\eqref{eq:ep_phonon}.

These self-energies are connected with the Fourier components of the Green's functions, see Eqs.~\eqref{eq:G_loc_def} and \eqref{eq:def_phonon_Green_Matsubara}.
We determine the self-energies self-consistently in the DMFT calculation.
Note that when we evaluate the susceptibilities, we add the corresponding external field part and compute the Green's functions of the low-symmetric states, as will be discussed in Sec.~\ref{sec:External_perturbations}.
In such states, we need to take into consideration the shift of the vibrational center of phonons as explained in Sec.~\ref{sec:shift_phonon}.

\subsection{External perturbations} \label{sec:External_perturbations}
\subsubsection{Susceptibilities against static fields}

Now, we explain the calculation method for the various susceptibilities.
In this paper, the susceptibility is directly evaluated through self-consistent calculations with an applied external field.

We consider the physical quantity $\mathscr O$ and its conjugate field $h$.
The corresponding Hamiltonian for the external field is given by
\begin{align}
    \mathscr H_{\rm ext} &= -h \mathscr O = \sum_i 
    \bm \Psi_i^\dg \check H_{\rm ext} \bm \Psi_i, 
\end{align}
where the site index $i$ is written explicitly to emphasize that the external field $h$ is spatially uniform.
Then, the susceptibility of $\mathscr O$ is defined as
\begin{align}
    \chi_{\mathscr O}
    &=
    \lim_{h\to 0} \frac{\partial \la \mathscr O \ra}{\partial h}
    \simeq 
    \frac{\la \mathscr{O}
    \ra
    |_{h} - \la \mathscr{O}
    \ra|_{h=0}}{h}, \label{eq:susceptibility}
\end{align}
where the average $\la \cdots \ra|_{h}$ is taken for a small but nonvanishing $h$, and $\la \cdots \ra|_{h = 0}$ 
for $h = 0$.
Namely, two self-consistent calculations need to be performed: one with a nonzero external field $h$, and the other with $h=0$.
Then, we evaluate the susceptibility following Eq.~\eqref{eq:susceptibility}.
Below, we list the specific forms of the physical quantities $\mathscr O$ 
for both the diagonal components $\mathscr O^{\eta \mu}$ (such as magnetism and orbital moments) and the off-diagonal components $\mathscr P^{\eta \mu}$ (pair amplitudes) in Nambu space.

\paragraph{Diagonal quantities}

The diagonal quantities ($c^\dg c$) are defined using the Gell-Mann and Pauli matrices as follows:
\begin{align}
  \mathscr{O}^{\eta\mu}
  &=
  \sum_{\gm\gm'}\sum_{\sg\sg'}
  c_{\gm\sg}^\dg 
  \lambda_{\gm\gm'}^\eta \sg_{\sg\sg'}^\mu 
  c_{\gm'\sg'},  \label{eq:diag_EF_moment}
\end{align}
which is specified by $\eta = 0, \cdots, 8$ (orbital degrees of freedom) and $\mu = 1,2,3$ (spin degrees of freedom).
The corresponding external field Hamiltonian is expressed as
\begin{align}
    \check{H}_{\mathrm{ext}}
    &=
    -h
    \begin{pmatrix}
        \widehat{\lambda^\eta \otimes \sg^\mu} & \hat{0} \\
        \hat{0} &  -( \widehat{\lambda^\eta \otimes \sg^\mu})^{\rm T}
    \end{pmatrix}
    . \label{eq:Hext_def_diag}
\end{align}
As mentioned in Sec.~\eqref{sec:model}, $\eta = 0$ represents charge degrees of freedom, $\eta = 2,5,7$ the magnetic orbital moment, and $\eta = 1,3,4,6,8$ the electric orbital moment.
In other words, the type of diagonal quantity and corresponding external field are uniquely determined by the choice of $\eta$ and $\mu$.

\paragraph{Off-diagonal quantities}
The off-diagonal quantities ($c^\dg c^\dg, cc$) are defined as
\begin{align}
  \mathscr{P}^{\eta\mu}
  &=
  \sum_{\gm\gm'}\sum_{\sg\sg'}
  c_{\gm\sg}^\dg
  \lambda_{\gm\gm'}^\eta (\sg^\mu\epsilon)_{\sg\sg'} 
  c_{\gm'\sg'}^\dg 
  + \mathrm{H. c. }, \label{eq:offd_EF_moment}
\end{align}
where $\epsilon=\imu\sg^y$ is an antisymmetric tensor.
The corresponding external field Hamiltonian becomes
\begin{align}
  \check{H}_{\mathrm{ext}}
  &=
  -h
  \begin{pmatrix}
       \hat{0} & \widehat{\lambda^\eta \otimes (\sg^\mu\epsilon)}\\
       -( \widehat{\lambda^\eta \otimes (\sg^\mu\epsilon)})^{\rm T} & \hat{0}
  \end{pmatrix}
  . 
\end{align}
Note that $\mu = 0$ represents the spin-singlet pair, and $\mu = 1,2,3$ represent the spin-triplet pairs.

\subsubsection{Odd-frequency ordering instabilities}

It is also interesting to study order parameters which depend on frequency.
The cases with an odd functional form are of particular interest, since they show qualitatively different behavior from the static order parameters in the preceding subsection.
The corresponding ordering instabilities are captured by the odd-frequency susceptibility \cite{Anders02, Anders02_2, Hoshino14, Hoshino17, Sakai04}.

To be more specific, let us consider a method to locally perturb the system, which is different from the previous subsection.
We assume the presence of a virtual fermionic bath, which is hybridized locally with the electrons.
This situation is described by the following Hamiltonian:
\begin{align}
    \mathscr H_{\rm ext}[f^\dg,f]
    &= \sum_{\al}\sum_{\gm\sg} (V_{\al,\gm \sg}  f^\dg_\al c_{\gm\sg} + W_{\al,\gm \sg} f^\dg_\al c^\dg_{\gm\sg} ) + {\rm H.c.}
    \nonumber \\
    &\ \ \ +
    \sum_{\al} E_\al f_\al^\dg f_\al, 
\end{align}
where $f_\al$ is the annihilation operator with quantum number $\al$ for the bath.
In the path-integral formalism, we can integrate out the fermionic bath without loss of generality, and the effective action is obtained as
\begin{align}
    \int \mathscr D \bar f \mathscr D f \, \exp \bigg\{ - \int \diff \tau ( \bar f\partial_\tau f + \mathscr H_{\rm ext}[\bar f,f]) \bigg\}
    = \epn^{- S_{\rm ext}}, 
\end{align}
where the effective action is described by the frequency-dependent external perturbation given by
\begin{align}
    S_{\rm ext} &= - \int \diff \tau \diff \tau' \ 
    \bar {\bm \Psi}(\tau) \check h (\tau-\tau') \bm \Psi(\tau')
    .
\end{align}
The time-dependent field $\check h(\tau)$ is described by $V$, $W$, and $E$.
If we take $h(\tau-\tau')\propto \delta(\tau-\tau')$, it reproduces the static external perturbation in the previous subsection.
On the other hand, in general, this time-dependent quantity can be an odd function with respect to the exchange of $\tau$ and $\tau'$, which corresponds to an odd-frequency perturbation in Fourier space.
While its specific realization is not trivial, here we have demonstrated that an odd-frequency perturbation is, in principle, physically possible by the coupling to a fermionic bath.

With the above considerations in mind, we introduce the odd-frequency susceptibility.
The physical quantity is given by
\begin{align}
    \la \mathscr{O}_{\mathrm{OF}}^{\eta\mu} \ra
    &=
    \sum_{\gm\gm'}\sum_{\sg\sg'}
    \lambda_{\gm\gm'}^\eta \sg_{\sg\sg'}^\mu 
    \frac{1}{\beta}\sum_n 
    g(\imu\omega_n)\hat{G}_{\gm'\sg',\gm\sg}(\imu\omega_n) , 
\end{align}
where the form factor may be chosen as $g(\imu\omega_n) = \imu \, \sgn\omega_n$ \cite{Hoshino11, Hoshino14, Hoshino15proc, Sakai04, Anders02, Anders02_2, Jarrell97}.
Correspondingly, the susceptibility is given by $\chi_{\mathrm{OF}}^{\eta\mu} = \la \mathscr{O}_{\mathrm{OF}}^{\eta\mu} \ra/h$ with $h$ defined by $h(\imu\omega_n) = h \ g(\imu\omega_n)$.
These quantities should be contrasted with Eqs.~\eqref{eq:diag_EF_moment} and \eqref{eq:offd_EF_moment},  which represent `even-frequency' susceptibilities.
A similar quantity can be considered for the pair amplitude, known as the odd-frequency pair.
Specifically, this can be expressed in terms of the Green's function by rewriting Eq.~\eqref{eq:offd_EF_moment} and introducing a time dependence.
In previous studies in Refs.~\cite{Belitz99, Solenov09, Kusunose11, Fominov15}, two types of relations have been discussed between the odd-frequency pair function and its conjugate quantity.
In this work, however, we assume the relation that corresponds to the usual Hermitian conjugate.

While previous studies have reported cases in which the odd-frequency susceptibility diverges \cite{Hoshino11, Hoshino14, Hoshino17}, there is no significant enhancement of the odd-frequency susceptibility within our perturbative calculations.
This may reflect the weakness of electron correlations in the lowest-order approximation.
Therefore, we focus solely on the response to a static external field.
Interestingly, even in this situation, odd-frequency pairs are induced and play a significant role in the magnetic susceptibility, as will be discussed in Sec.~\ref{sec:susceptibilities}.

\section{Numerical results \label{sec:results}}

\subsection{Physical quantities considered in the analysis}

In the following, we list the specific physical quantities whose susceptibilities are investigated in this paper.
We begin with the $z$-component of the spin magnetic moment, corresponding to the case $(\eta, \mu) = (0, 3)$ in Eq.~\eqref{eq:diag_EF_moment}:
\begin{align}
  S^z
  &=
  \sum_\gm \sum_{\sg\sg'}
  c_{\gm\sg}^\dg \sg_{\sg\sg'}^3 c_{\gm\sg'}. 
\end{align}
Since the $x$- and $y$-components of the magnetic moment can be obtained by rotating the quantization axis, we focus only on the $z$-component in the following. 
We also consider the quantities related to the orbital degrees of freedom, namely $\eta \neq 0, \mu = 0$.
Here, we list the $z$-component of the magnetic orbital moment and the $x^2 - y^2$-component of the electric orbital moment:
\begin{align}
  L^z
  &=
  \sum_{\gm\gm'} \sum_\sg
  c_{\gm\sg}^\dg \lambda_{\gm\gm'}^2 c_{\gm'\sg}
  , \label{eq:Lz}\\
  Q^{x^2-y^2}
  &=
  \sum_{\gm\gm'} \sum_\sg
  c_{\gm\sg}^\dg \lambda_{\gm\gm'}^3 c_{\gm'\sg}
  , \label{eq:Q_x2-y2}
\end{align}
which are obtained by setting $(\eta, \mu) = (2, 0)$ and $(\eta, \mu) = (3, 0)$ in Eq.~\eqref{eq:diag_EF_moment}, respectively.
The physical interpretations of Eqs.~\eqref{eq:Lz} and \eqref{eq:Q_x2-y2} are as follows: the orbital magnetization in $L^z$ corresponds to the magnetization originating from the orbital motion of the electrons rotating around the the $z$-axis (see upper right panel of Fig.~\ref{fig:Gell-Mann}), while the quadrupole $Q^{x^2 - y^2}$ represents the distortion of the charge distribution of the $x^2 - y^2$ type, 
reflecting the orbital degrees of freedom (see lower panel of Fig.~\ref{fig:Gell-Mann}).

As an off-diagonal quantity, we consider the intra-orbital spin-singlet ($s$-wave) pair amplitude with $(\eta, \mu) = (0, 0)$:
\begin{align}
  p^{r^2}
  &=
  \sum_{\gm\gm'}\sum_{\sg\sg'}
  c_{\gm\sg}^\dg
  \lambda_{\gm\gm'}^0 (\sg^0\epsilon)_{\sg\sg'} 
  c_{\gm'\sg'}^\dg 
  +{\rm H.c.} . 
\end{align}
This is the order parameter of the superconducting phase realized as an ordered state, which is consistent with other theoretical \cite{Nomura15_2, Nomura16, Akashi13} and experimental \cite{Zhang91, Zhang91_2, Tycko92, Rotter92, Auban-Senzier93, Kiefl93, Sasaki94, Degiorgi94, Gu94, Koller96, Hesper00} results for fulleride superconductors.

\subsection{Model parameters relevant to fullerides}

As mentioned in Sec.~\ref{sec:model}, we investigate a model based on the single-mode approximation, representing the phonon modes by one $A_g$ mode and one $H_g$ mode.
To perform the calculations, it is necessary to specify the parameters. 
To parametrize the electron-phonon coupling strength, we introduce the following coupling constants for the effective phonon-mediated interaction:
\begin{align}
    \lambda_\eta^{\rm st} = \frac{4g_\eta^2}{3\omega_\eta}, \label{eq:lambda_coupling}
\end{align}
where `st' indicates the static part of the electron-phonon interaction. 
Hence, we need to choose the following parameters: the bandwidth $D$, phonon frequencies $\omega_0$ and $\omega_1 \ (= \omega_{3,4,6,8} )$, effective interactions $\lambda^{\rm st}_0$ and $\lambda^{\rm st}_1 \ (= \lambda^{\rm st}_{3,4,6,8} )$ [Eq.~\eqref{eq:lambda_coupling}], and Coulomb interactions $U$ and $J$. 
We determine these parameters 
so as to roughly reproduce the experimentally observed transition temperature, 
employing 
the effective phonon-mediated interactions obtained from first-principles calculations \cite{Faber11, Nomura15, Nomura15_2}. 
We furthermore use the notation 
$I_\eta^{\rm ep} = -3\lambda^{\rm st}_\eta/4$ with $I_0^{\mathrm{ep}}=3U^{\mathrm{ep}}/4-J^{\mathrm{ep}}$ and $I_{1,3,4,6,8}^{\mathrm{ep}}=J^{\mathrm{ep}}/2$, which corresponds to the $I_\eta$ parameters in the Coulomb interaction term [Eq.~\eqref{eq:ham_int_I}].

The dressed phonon may be estimated 
as $\tilde{\omega}_\eta \simeq 0.18$ eV \cite{Nomura15, Faber11, Nomura15_2}, which corresponds to the resonance frequency of the 
effective interaction.
The relation between the bare and renormalized phonon frequency is given by \cite{Kaga22}
\begin{align}
    \omega_\eta&= \tilde{\omega}_\eta \left(1 + \frac{8}{\pi D}\tilde{\lambda}^{\rm st}_\eta\right).  
    \label{eq:tilde_omega}
\end{align}
We have introduced $ \tilde \lambda_\eta ^{\rm st} = 4g_\eta^2 / 3\tilde \omega_\eta $.
The phonon frequency appearing in $I_\eta^{\rm ep}$ is taken to be $\omega_\eta$ as defined in Eq.~\eqref{eq:tilde_omega}. 

\begin{figure*}[tb]
    \centering
    \includegraphics[width=180mm]{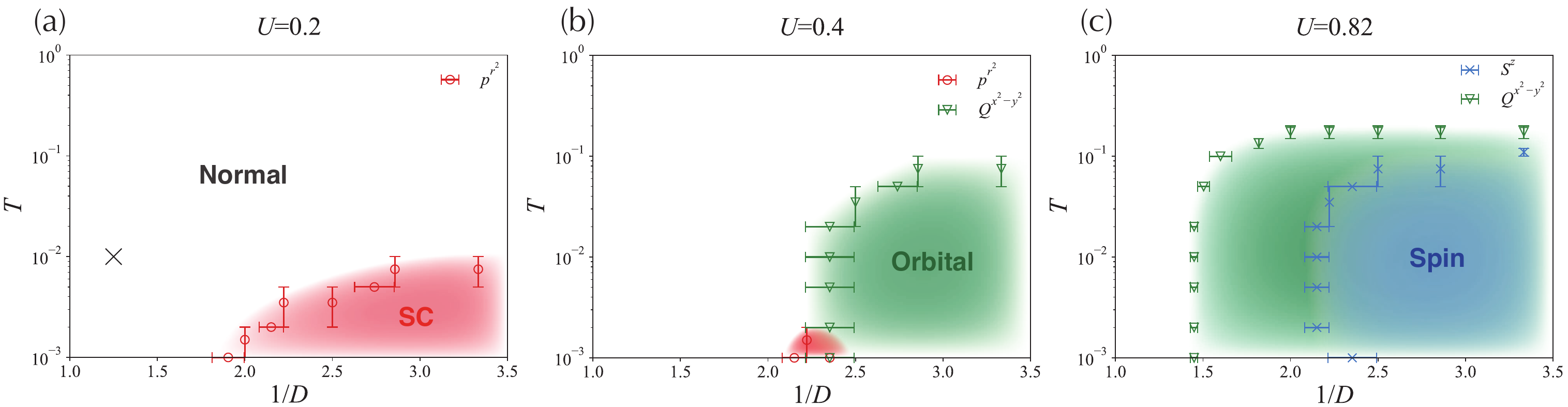}
    \caption{
    Phase diagrams of the Jahn-Teller Hubbard model plotted as a function of the inverse bandwidth $1/D$ and temperature $T$. Each panel corresponds to a different value of the Coulomb interaction: (a) $U=0.2$ eV, (b) $U=0.4$ eV, and (c) $U=0.82$ eV.
    The black cross symbol in (a) corresponds to the parameter which we use in Sec.~\ref{sec:lam1_vs_J} and Fig.~\ref{fig:lam1_vs_J}.}
    \label{fig:phase_diagram}
\end{figure*}

Using the results in Ref.~\cite{Nomura16}, we estimate $U^{\mathrm{ep}}=-0.152\,\mathrm{eV}$, $J^{\mathrm{ep}}=-0.050\,\mathrm{eV}$, from which we determine the bare parameters. 
Inserting these values into the expressions of $I_\eta^{\rm ep}$  
and using Eq.~\eqref{eq:tilde_omega}, we obtain $\lambda^{\rm st}_0=0.085$ eV, $\lambda^{\rm st}_{1,3,4,6,8}=\lambda^{\rm st}_1=0.033$ eV, $\omega_0=0.318\,\mathrm{eV}$, $\omega_{1,3,4,6,8}=\omega_1=0.217\,\mathrm{eV}$.
Whereas these parameters give qualitatively consistent results for alkali-doped fullerides in the weak-coupling regime, a description of the multi-mode phonons instead of the single-mode approximation is necessary for a more quantitative description.

For the Coulomb interaction part, we use the 
values $U=0.82\,\mathrm{eV}$, $J=0.031\,\mathrm{eV}$, which are based on constrained random phase approximation (cRPA) calculations \cite{Nomura12, Nomura16}.
We take $\mu = 0$, which corresponds to half-filling for the three-orbital model, and is relevant for the alkali-doped fullerides A$_3$C$_{60}$.
The test field in the susceptibility calculation is chosen as $h=0.001$ eV.

\subsection{Phase diagram}

Let us map out the phase diagram in the plane of temperature and bandwidth.
Experimentally, an increase of the bandwidth corresponds to the application of pressure, which changes the spatial distance between the fullerene molecules.

First of all, we provide a rough physical picture for the ordered states in our system.
We have two energy scales, i.e., the effective attractive interaction $\lambda^*$, which leads to Cooper pair formation (superconductivity), and the effective repulsive interaction $U^*$, which leads to spin or orbital ordered states.
Because of the Fermi surface instability, the infinitesimal attractive force drives the system to the superconducting (SC) state at sufficiently low temperatures.
The transition temperature is roughly estimated as
\begin{align}
    T_{\rm SC} &\sim \omega_c \exp \qty( - \frac{\lambda^*}{D} ), 
\end{align}
where $D$ is the bandwidth (the density of states at the Fermi level is $\sim 1/D$), and $\omega_c$ is a cutoff frequency for the effectively attractive interaction.
This is identical to the BCS formula for conventional superconductors.
On the other hand, if the repulsive interaction is sufficiently large, the ground state should become a spin (and/or orbital) ordered state.
The condition for the appearance of this ordered states is 
\begin{align}
    \frac{U^*}{D} \gtrsim 1.
\end{align}
This qualitatively corresponds to the Stoner mechanism.
Thus, the ground state for $U^* \gtrsim D$ is a diagonally ordered state, while for $U^* \lesssim D$  superconductivity appears at low $T$.

Since the bare Coulomb repulsive parameter $U$ is so large that the superconductivity cannot appear, we consider $U$ as an effectively renormalized parameter.
Indeed, higher-order corrections screen the Coulomb interaction, and the effective magnitude should be smaller for the dressed interaction.
Since our perturbative Eliashberg approach includes only the lowest-order corrections, the screening from the higher-order terms is included in the form of a reduced $U$ parameter.

Figure~\ref{fig:phase_diagram} shows the phase diagram in the plane of $1/D$ and $T$ for 
$U=\{0.2 \ {\rm eV},0.4 \ {\rm eV},0.82 \ {\rm eV}\}$. 
We set the parameters as $\lambda^{\rm st}_0=0.085$ eV, $\lambda^{\rm st}_1=0.033$ eV, $\omega_0=0.318$ eV, $\omega_1=0.217$ eV, $J=0.031$ eV.
The case with $U=0.2$ eV is shown in Fig.~\ref{fig:phase_diagram}(a). 
The $s$-wave superconducting phase appears when the inverse bandwidth becomes $1/D \gtrsim 1.9 \, \mathrm{eV^{-1}}$ (i.e. $D \lesssim 0.53$ eV).
As the bandwidth decreases, the superconducting transition temperature increases.
In the case of $U = 0.4$ eV shown in Fig.~\ref{fig:phase_diagram}(b), the superconducting phase is suppressed and electric-orbital order 
is realized at large $1/D \gtrsim 2.4\, \mathrm{eV^{-1}}$.
When the Coulomb interaction is further increased to the bare value $U=0.82$ eV [Fig.~\ref{fig:phase_diagram} (c)], an electric orbital order emerges around $1/D = 1.5\, \mathrm{eV^{-1}}$.
As the bandwidth is reduced further, a spin magnetization also appears around $1/D=2.2 \, \mathrm{eV^{-1}}$.

\begin{figure*}[tb]
    \centering
    \includegraphics[width=170mm]{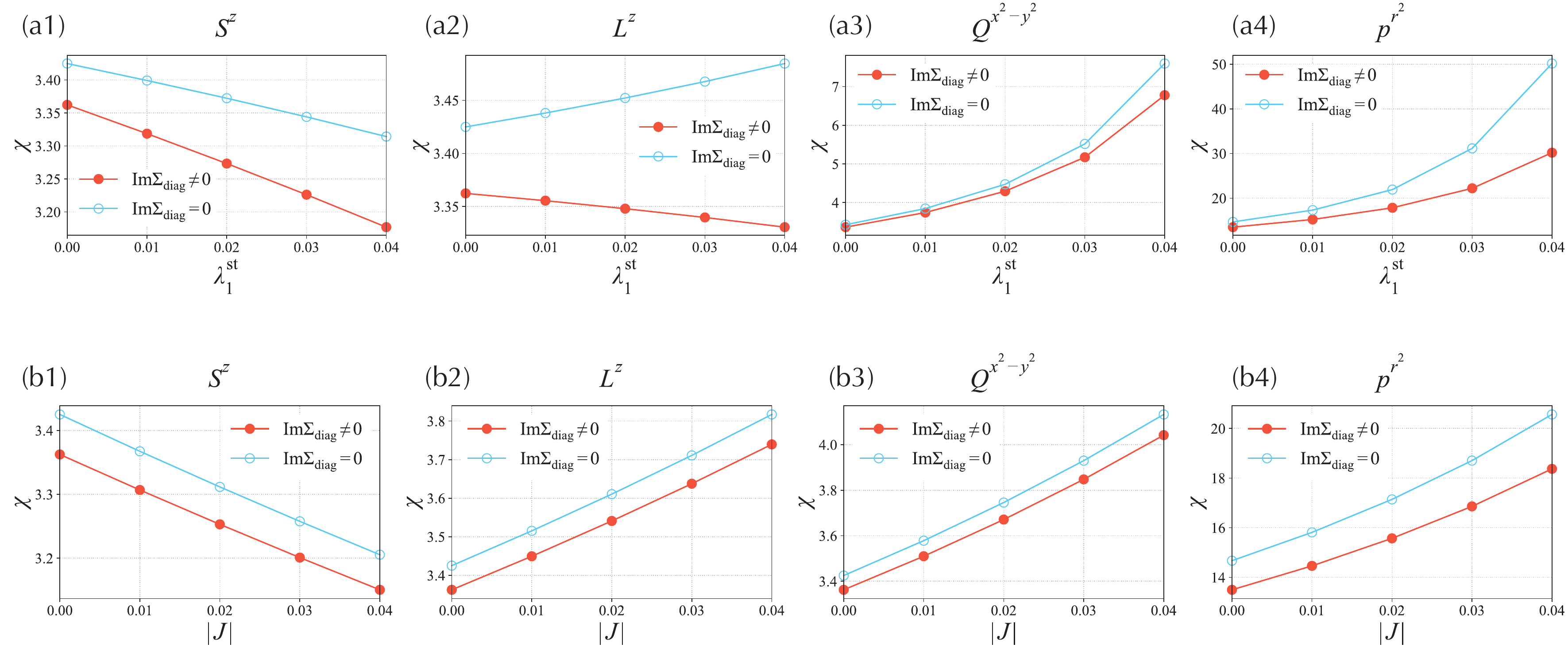}
    \caption{
    Comparison of the susceptibility behavior as a function of electron-phonon coupling $\lambda^{\rm st}_1$ (upper panels, (a1)–(a4)) and Hund's coupling $J$ (lower panels, (b1)–(b4)). 
    The columns, from left to right, show the susceptibilities of the magnetic moment, orbital magnetic moment, quadrupole, and spin-singlet pair, respectively. 
    }
    \label{fig:lam1_vs_J}
\end{figure*}

In actual fullerides, spin and orbital orderings do not appear in the weak correlation regime.
Thus, among the three phase diagrams in Fig.~\ref{fig:phase_diagram}(a)-(c), the one at $U=0.2$ eV is the closest to what is realized in actual materials.
This small value of $U$ can be roughly understood by considering screening effects within the random phase approximation (RPA), 
which can be roughly estimated as
\cite{Kaga22}
\begin{align}
    \tilde I_\eta = \frac{I_\eta}{1 + 2I_\eta \chi}
    .
\end{align}
We can evaluate the local charge susceptibility $\chi$ from a simple bubble-type diagram as $\chi \simeq 32 / 3\pi D$ \cite{Kaga22}. 
With the bare parameter $U=0.82$ eV and bandwidth $D=0.25$ eV, the screening factor is 
estimated as $(1+2I_0 \chi)^{-1} \simeq 0.056$.
Hence, the effective Coulomb interaction parameter $U$ can be approximately ten times smaller than the bare parameter.
Note that the anisotropic part, the Hund's coupling $J$, is not strongly renormalized because it is a non-density-type interaction and does not directly couple to the total charge.

\subsection{Coupling to Jahn-Teller phonons (\texorpdfstring{$\lambda_1^{\rm st}$}{lambda1}) versus effective antiferromagnetic Hund's coupling (\texorpdfstring{$J<0$}{J}) \label{sec:lam1_vs_J}}

In this subsection, we elucidate the contributions of the Jahn-Teller phonons to 
the response functions
above the superconducting transition temperature. 
We analyze the qualitative behaviors of
 the spin magnetization $S^z$, orbital magnetization $L^z$, orbital quadrupoles $Q^{x^2-y^2}$, and spin-singlet pair amplitudes $p^{r^2}$.
We take the parameters $U=0.2$ eV and $D=0.8$ eV, 
which avoid the spin and orbital ordered phases as discussed in the previous subsection.
We set the temperature to $T = 0.01$ eV, which corresponds to the normal state, as shown by the black cross symbol in Fig.~\ref{fig:phase_diagram}(a).

To clarify the retardation effects induced by the coupling to Jahn-Teller phonons, we consider two cases: on the one hand, we study the susceptibilities as a function of the phonon-mediated electron interaction $\lambda_1^{\rm st}$ with $J = 0$.
On the other hand, we consider the antiferromagnetic Hund's coupling $J<0$ with $\lambda^{\rm st}_1 = 0$, so that the retardation effect of the Jahn-Teller phonon coupling is neglected.
This allows us to investigate how the presence ($J = 0, \lambda^{\rm st}_1 > 0$) or absence ($J < 0, \lambda^{\rm st}_1 = 0$) of the retardation effect alters the behavior of the susceptibilities.

The orange lines (filled symbols) in Figs.~\ref{fig:lam1_vs_J}(a1)–(a4) show the $\lambda^{\rm st}_1$ dependence of the susceptibilities for the spin magnetization $S^z$, the orbital magnetization $L^z$, the quadrupole $Q^{x^2 - y^2}$, and the $s$-wave superconductivity $p^{r^2}$. 
The susceptibilities increase with increasing $\lambda^{\rm st}_1$ for the quadrupole $Q^{x^2 - y^2}$ in Fig.~\ref{fig:lam1_vs_J}(a3) and the $s$-wave superconducting $p^{r^2}$ channel in Fig.~\ref{fig:lam1_vs_J}(a4).
In contrast, the susceptibilities for the spin magnetization $S^z$ in Fig.~\ref{fig:lam1_vs_J} (a1) and orbital magnetization $L^z$ in Fig.~\ref{fig:lam1_vs_J}(a2) decrease as $\lambda^{\rm st}_1$ increases.

On the other hand, the 
orange lines with filled circles in Figs.~\ref{fig:lam1_vs_J}(b1)–(b4) show the $|J|$-dependence of the susceptibilities for $\lambda^{\rm st}_1 = 0$. 
The susceptibilities increase with increasing $|J|$ for the orbital magnetization [Fig.~\ref{fig:lam1_vs_J}(b2)], quadrupole [Fig.~\ref{fig:lam1_vs_J}(b3)], and $p^{r^2}$ [Fig.~\ref{fig:lam1_vs_J}(b4)].
Only the spin magnetization in Fig.~\ref{fig:lam1_vs_J}(b1) shows a decreasing susceptibility with increasing $|J|$.

Comparing the $\lambda^{\rm st}_1$-dependence of the susceptibility [orange lines of Fig.~\ref{fig:lam1_vs_J}(a1)-(a4)] with its $J$-dependence [orange lines of Fig.~\ref{fig:lam1_vs_J}(b1)-(b4)], 
only the orbital magnetization $L^z$ shows a qualitative difference in behavior.
Namely, while it decreases with respect to $\lambda^{\rm st}_1$, it increases with respect to $|J|$.
Therefore, we can conclude that the behavior of the orbital magnetization changes qualitatively
if the retardation effect of the Jahn-Teller phonons is switched off.

In order to clarify the origin of the observed behavior, i.e., that 
the orbital magnetization $L^z$ is strongly affected by retardation effects, we focus on $\imag\hat{\Sigma}_{\mathrm{diag}}(\imu\omega_n)$, which has a finite contribution only when the retardation effect is present. 
The orbital susceptibility calculated by setting the obtained diagonal self-energy to $\imag\hat{\Sigma}_{\mathrm{diag}}(\imu\omega_n) = 0$ is shown by the light-blue line (open symbols) in Fig.~\ref{fig:lam1_vs_J}.
By comparing the cases with (orange lines) and without (light-blue lines) dynamic effects, we 
find that only in the case of the orbital magnetization does the susceptibility exhibit a qualitatively different behavior [Fig.~\ref{fig:lam1_vs_J}(a2)].
This clearly indicates that the retardation effect of the coupling to phonons, particularly the contribution from $\imag\hat{\Sigma}_{\mathrm{diag}}(\imu\omega_n)$, plays a significant role in reducing the susceptibility.

\section{Magnetic spin and orbital susceptibilities \label{sec:susceptibilities}}

\subsection{Temperature dependence \label{sec:tmperature}}

\begin{figure}[tb]
    \centering
    \includegraphics[width=85mm]{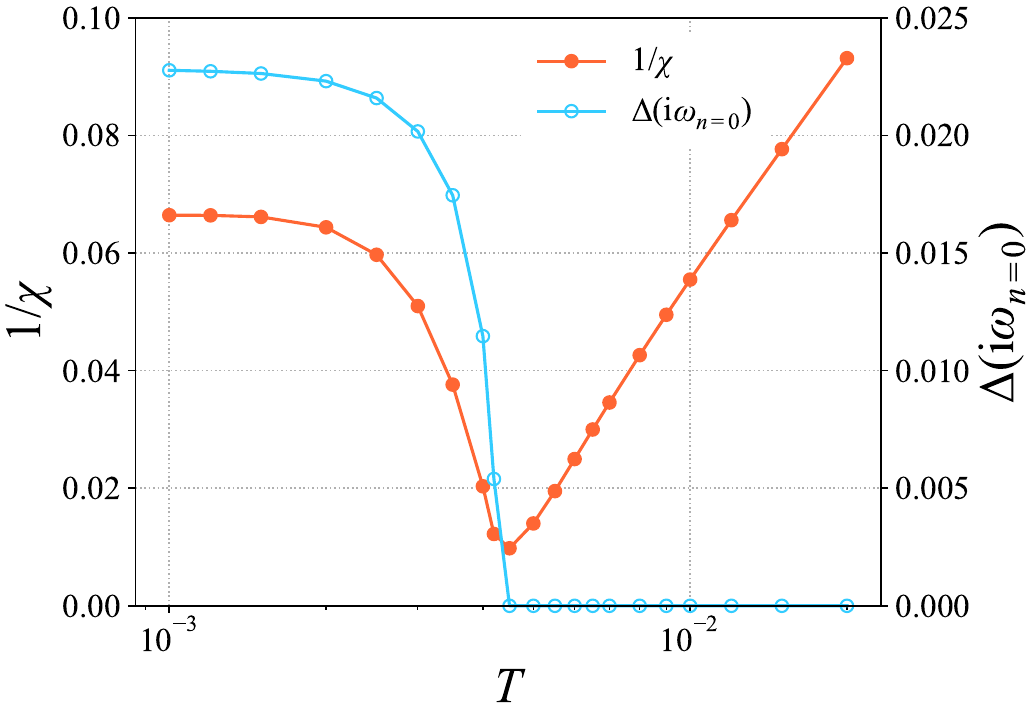}  
    \caption{
    Temperature dependence of the inverse susceptibility for the spin-singlet pairing (filled symbols) and 
    the order parameter 
    $\Delta(\imu\omega_{n=0})$ without an external field (open symbols). 
    }
    \label{fig:singlet_Tdep}
\end{figure}

In the previous section, we analyzed the susceptibility in the normal state (metallic phase) and compared the cases with and without retardation effects. 
In this section, we investigate the retardation effect in the superconducting state from the perspective of the temperature dependence of the susceptibility.
For the coupling constant of the electron-phonon interaction and phonon frequencies, we use those obtained in the previous section. 
The Coulomb interaction and the bandwidth are set to $U = 0.2$ eV and $2D = 0.8\,\mathrm{eV}$, respectively.

\begin{figure}[tb]
    \centering
    \includegraphics[width=85mm]{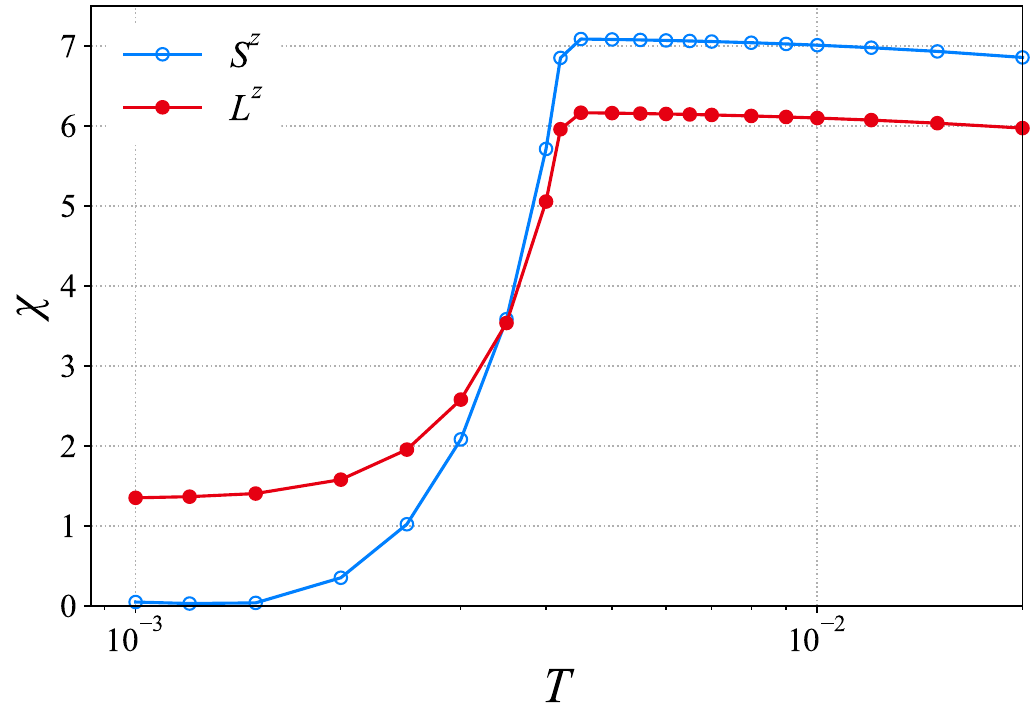}  
    \caption{
    Temperature dependence of the magnetic susceptibility (open symbols) and orbital magnetic susceptibility (filled symbols).
    }
    \label{fig:Sz-Lz_Tdep} 
\end{figure}

We begin by examining the pair potential (anomalous self-energy at the lowest Matsubara frequency) and the superconducting susceptibility.
Fig.~\ref{fig:singlet_Tdep} shows the temperature dependence of the inverse susceptibility for spin-singlet pairing (orange
line, filled symbols) and the pair potential (blue line, open symbols).
The inverse susceptibility takes a minimum value at $T \simeq 0.0045$ eV, indicating the transition temperature, and then gradually increases as the temperature decreases further.
Note that the inverse susceptibility does not reach zero exactly at the transition temperature due to the finite magnitude of the external field. 
Correspondingly, the pair potential
becomes nonzero below the transition temperature.

Now, we move on to the results for the spin and orbital magnetic susceptibilities.
Figure~\ref{fig:Sz-Lz_Tdep} shows the temperature dependence of the spin (blue line, open symbols) and orbital magnetic (red line, filled symbols) susceptibilities.
The spin susceptibility (blue line) approaches zero below the transition temperature, with a functional form similar to the Yosida function in BCS theory \cite{Yosida58}.
On the other hand, although the orbital magnetic susceptibility decreases below the transition temperature, it remains nonzero at low temperatures (red line).
It is intriguing that a nonvanishing orbital magnetic susceptibility exists in the superconducting state, even though we consider an $s$-wave spin-singlet superconductor.

The suppression of the spin susceptibility at low temperatures can be intuitively understood by the spin-singlet nature of the Cooper pairs, which renders the spin Zeeman effect inactive. 
Regarding the orbital Zeeman effect, the presence of orbital degrees of freedom can explain a nonzero susceptibility, since the 
orbital part of
the Cooper pairs 
has no singlet-like structure
(i.e., is orbital-symmetric). Hence, one can naively think of a  circulating motion of the pair within the molecular orbitals.
However,
as pointed out in our previous work \cite{Okada_proc}, the orbital susceptibility vanishes in the absence of retardation effects. 
Furthermore, as will be shown below, when retardation is present, the pair potential at high frequencies 
becomes suppressed compared to the case without retardation, which could 
lead one to expect a finite spin susceptibility.

It is therefore necessary to identify the origin of the behavior observed in this subsection, which will be discussed in detail in the next subsection.
By simplifying the model and method without affecting the qualitative results, we find that odd-frequency pairing plays a crucial role in ensuring that the spin susceptibility vanishes at low temperatures.
Similarly, the nonvanishing orbital magnetic susceptibility can be understood as a consequence of a smaller amount of odd-frequency pairing induced by the orbital Zeeman effect, compared to the spin Zeeman effect. This is due to the multiorbital nature, which manifests itself in the non-commutativity of the Gell-Mann matrices.

\subsection{Linear response theory for a simplified model}

An interesting observation in the previous subsection is that the magnetic susceptibility for the orbital angular momentum remains nonzero at low temperatures, 
while the spin magnetic susceptibility goes to zero for $T\to 0$. 
Since both the orbital and spin magnetic moments contribute to the physically observable total magnetic moment \cite{Tosatti96}, it is important to clarify the origin of those susceptibility behaviors.

The discussion in this subsection proceeds as follows.
In Sec.~\ref{sec:review_bcs}, the vanishing spin susceptibility in BCS theory is reviewed.
We formulate the linear response theory for the Eliashberg equation in Sec.~\ref{sec:linear_response}, while the numerical results are presented in Secs.~\ref{sec:numerical_tmp} and \ref{sec:numerical_freq}.
The analytic solution is derived in Sec.~\ref{sec:strong_ret} for the strong retardation limit to further support the intepretation 
in the preceding subsection.
We also discuss alternative physical quantities for the odd-frequency pair amplitudes in Sec.~\ref{sec:TimeIndep}.

\subsubsection{Brief review of spin susceptibility in BCS theory \label{sec:review_bcs}}

While the above anomalous behavior is caused by the retardation effect and multiorbital nature, 
it is useful to review the zero spin susceptibility at low $T$ within the conventional BCS theory without retardation effects.

In the standard BCS theory, 
once the frequency-independent static) gap parameter $\Delta_{\rm st}
>0$ has been obtained by solving the BCS gap equation, the linear response to the external spin magnetic field $h$ gives the following magnetic moment $M$ \cite{Okada_proc}:
\begin{align}
    &M = 4T\sum_n \int \diff \ep \rho(\ep)G_1(\ep,\imu\omega_n)
    , \label{eq:Mdef_BCS}
    \\
    &G_1(\ep,\imu\omega_n)/h = -  \mathsf g_0(\ep,\imu\omega_n)^2 - \mathsf f_0(\ep,\imu\omega_n)^2  
    ,
    \label{eq:BCS_bubble}
\end{align}
where $G_1$ is the $O(h^1)$ deviation of the Green's function by the external field relevant to the spin Zeeman effect.
$\mathsf g_0$ and $\mathsf f_0$ are the normal and anomalous Green's functions, respectively, given by
\begin{align}
    \mathsf g_0(\ep,\imu\omega_n)
    &=  \frac{\imu\omega_n + \ep}{(\imu\omega_n)^2-\ep^2 -\Delta_{\rm st}^2}
    ,
    \\
    \mathsf f_0(\ep,\imu\omega_n)
    &=  \frac{\Delta_{\rm st}}{(\imu\omega_n)^2-\ep^2 -\Delta_{\rm st}^2}
    .
\end{align}
Diagrammatically, Eq.~\eqref{eq:BCS_bubble} corresponds to a bubble contribution.
Note that the factor of 4 in Eq.~\eqref{eq:Mdef_BCS} originates from spin-orbital degrees of freedom.

We proceed with the evaluation of the susceptibility in two ways: Performing (i) the Matsubara-sum first or (ii) the $\ep$-integral first.
More specifically, we define the following two auxiliary functions $G_1^{\rm (i),(ii)}$ which serve as a spectral decomposition for the magnetic susceptibility:
\begin{align}
    \chi_{\rm BCS}( \Delta_{\rm st}) \equiv \frac{M}{h} &=
    \frac{4}{h}\int \diff \ep \rho(\ep) G_1^{\rm (i)}(\ep; \Delta_{\rm st})
    \label{eq:chi_stat_epsil}
    \\
    &=
    \frac{4T}{h} \sum_n G_1^{\rm (ii)}(\imu\omega_n; \Delta_{\rm st})
    \label{eq:chi_stat_matsu}
    .
\end{align}
Without retardation effects, we can explicitly derive both expressions.
For case (i), we obtain 
\begin{align}
        G_1^{\rm (i)}(\ep; \Delta_{\rm st})/h &= - \frac{\partial f_{\rm F}(E)}{\partial E}, 
\end{align}
where $E(\ep) = \sqrt{\ep^2 + \Delta_{\rm st}^2}>0$.
$f_{\rm F}(x) = 1/(\epn^{\beta x}+1)$ is the Fermi distribution function.
Since the derivative of the Fermi distribution function has a nonzero value for $|x| \lesssim T$,
the spin susceptibility vanishes for $T\ll \Delta_{\rm st} < E(\ep)$.
The resulting temperature dependence of $\chi_{\rm BCS}$ is known as Yosida function \cite{Yosida58}.

\begin{figure}[tb]
    \includegraphics[width=85mm]{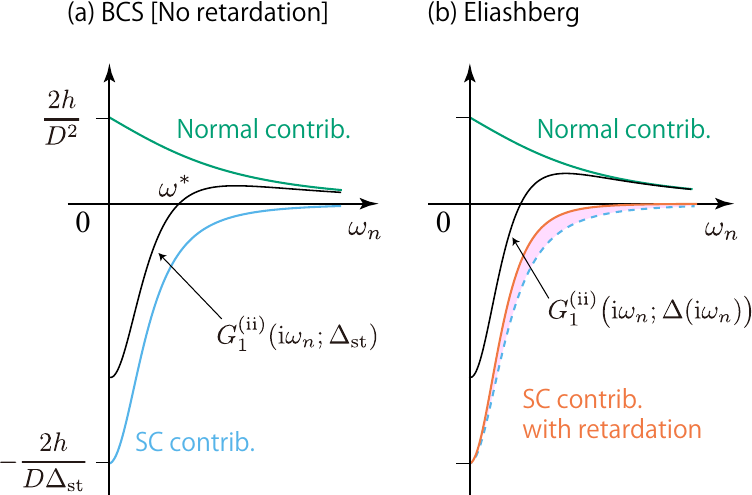}  
    \caption{
Schematic illustrations for the spectral decomposition of the magnetic susceptibility defined in Eq.~\eqref{eq:chi_stat_matsu}. Static and retarded pair potentials are used in (a) and (b), respectively, where the functional form $\Delta(\imu\omega_n) = \Delta_{\rm st} \frac{\omega_{\rm ph}^2}{\omega_n^2 + \omega_{\rm ph}^2}$ is employed for the plot in (b).
The shaded region (pink color) in panel (b) indicates a missing contribution due to the pair potential with retardation effect. \label{fig:Schematic_BCS}} 
\end{figure}

For the case (ii), on the other hand, we first perform the $\ep$-integration. The advantage of this approach is its applicability to cases where the self-energies depend on the Matsubara frequency (see Sec.~\ref{sec:numerical_freq}). 
Also, this approach has enabled us to analyze the Matbubara frequency dependence, resulting in an elucidation of nontrivial odd-frequency pairing contribution as discussed in the following subsections. 
In the present case, 
we obtain
\begin{align}
    G_1^{\rm (ii)}(\imu\omega_n; \Delta_{\rm st})/h &=  \Lambda'(\imu\Omega_n) + \qty( - \Lambda'(\imu\Omega_n) + \frac{\Lambda(\imu\Omega_n)}{\imu\Omega_n} ) \frac{\Delta_{\rm st}^2}{\Omega_n^2} 
    \label{eq:G1_type_ii}
    ,
\end{align}
where we have defined $\Omega_n = \sqrt{\omega_n^2 + \Delta_{\rm st}^2}>0$. 
We have also introduced the Hilbert transformation $\rho \to \Lambda$ for the complex variable $z=\imu\Omega_n$ as
\begin{align}
    \Lambda(\imu\Omega_n) &\equiv \int \diff \ep \frac{\rho(\ep)}{\imu\Omega_n-\ep}
    = \frac{2\imu(\Omega_n-  \sqrt{\Omega_n^2+D^2})}{D^2}
    \label{eq:dos_analytical}
    .
\end{align}
Now we assume that the band width is much larger than the gap function, so that  Eq.~\eqref{eq:G1_type_ii} simplifies to 
\begin{align}
    \frac{G_1^{\rm (ii)}(\imu\omega_n; \Delta_{\rm st})}{h} &\simeq \frac{2(\sqrt{\omega_n^2+D^2}-|\omega_n|)}{D^2\sqrt{\omega_n^2+D^2}}  - \frac{2 \Delta_{\rm st}^2}{D(\omega_n^2+\Delta_{\rm st}^2)^{3/2}}.
    \label{eq:G1_BCS_case_ii}
\end{align}
The first term of Eq.~\eqref{eq:G1_BCS_case_ii} is a contribution of the normal part, which is counteracted by the superconducting contribution in the second term with the opposite sign.
It is also notable that the superconducting contribution exists dominantly for $\omega_n \lesssim \Delta_{\rm st}$, while the normal contribution remains up to the energy scale of the bandwidth.
Thus, from the viewpoint of the Matsubara frequency dependence, the high-energy normal contribution (positive) is counteracted by the superconducting one (negative) at low frequencies, to result in the vanishing magnetic spin-Zeeman susceptibility.
The characteristic frequency $\omega^*$ where the contribution changes from positive to negative is estimated as $\omega^* \simeq (D\Delta_{\rm st}^2)^{1/3}$ where $\Delta_{\rm st}\ll\omega^* \ll D$ is satisfied.
The situation is schematically illustrated in Fig.~\ref{fig:Schematic_BCS}(a).

In the presence of the retardation effect, the gap function is replaced by $\Delta_{\rm st} \to \Delta (\imu\omega_n)$, which has a dominant contribution to the gap function only for $\omega_n \lesssim \omega_{\rm ph}$, where $\omega_{\rm ph}$ is a characteristic phonon energy scale.
Then, we recognize the missing contribution for $\omega_n \gtrsim \omega_{\rm ph}$ compared to Eq.~\eqref{eq:G1_BCS_case_ii}, resulting in the incomplete cancellation and hence a nonvanishing spin susceptibility.
We illustrate the frequency dependence of 
this superconducting contributions in Fig.~\ref{fig:Schematic_BCS}(b).
This argument indicates that, in order to obtain a vanishing susceptibility, there must be an additional contribution in the presence of retardation, which is discussed in detail below.

\subsubsection{Theoretical framework of linear response theory \label{sec:linear_response}}

Now let us consider the retardation effect and multiorbital nature in order to account for the vanishing susceptibility for the spin and the nonzero susceptibility for the orbital magnetic moment observed in Sec.~\ref{sec:tmperature}.
We employ the linear response theory within the DMFT+Eliashberg framework, and expand the Green's function with respect to the external magnetic field as
\begin{align}
    \check {G}(\imu\omega_n) &= \int \diff \ep \rho(\ep)
    \qty[ \imu\omega_n \check {1}- \ep \check {\tau}_3 - \check {H}_{\rm ext} - \check {\Sigma}(\imu\omega_n) ]^{-1}
    \label{eq:g_ext_gen}
    \\
    &= \check {G}_0(\imu\omega_n) + \check {G}_1(\imu\omega_n) + O(H_{\rm ext}^2).
    \label{eq:g_ext_expand}
\end{align}
where $\check H_{\rm ext}$ is given in Eq.~\eqref{eq:Hext_def_diag}. 
The self-energy is also expanded in powers of the external field as $\check {\Sigma} = \check {\Sigma}_0 + \check {\Sigma}_1 + \cdots$, where
\begin{align}
    \check {\Sigma}_0(\tau) &= - \sum_{\eta\eta'} g_\eta g_{\eta'} D_{\eta\eta'}(\tau)
    \check {\lambda}^{\eta} \check {G_0}(\tau) \check {\lambda}^{\eta'},
    \label{eq:perturb0}
    \\
    \check {\Sigma}_1(\tau) &= - \sum_{\eta\eta'} g_\eta g_{\eta'} D_{\eta\eta'}(\tau)
    \check {\lambda}^{\eta} \check {G_1}(\tau) \check {\lambda}^{\eta'}.
    \label{eq:perturb1}
\end{align}
Note that the equation for the self-energy is written in terms of imaginary time, and the Green's function in Matsubara frequency.
This property makes the analytic computation difficult.
Hence we take both numerical and analytic approaches, the latter of which is performed in the strong retardation limit and is discussed in Sec.~\ref{sec:strong_ret}.

 In this subsection, we are interested in the qualitative aspects of the susceptibility behaviors.
Since the presence or absence of the Coulomb interaction does not affect the qualitative behavior, it is dropped from Eqs.~\eqref{eq:perturb0} and \eqref{eq:perturb1}; only the electron-phonon interaction is considered here.
In addition, although the phonon self-energy is influenced by the external field, its contribution to the electronic self-energy is $O(g_\eta^4)$.
This higher-order contribution is neglected here, i.e., we keep only the leading-order $O(g_\eta^2)$ contribution to the electron self-energy.
This approximation of neglecting the phonon self-energy is  justified by the fact that it does not change the qualitative behaviors of the susceptibilities, as confirmed by numerical calculations without phonon self-energy.

We consider the spin-singlet pairing state.
The zeroth-order self-energy has the form 
\begin{align}
    \check {\Sigma}_0(\imu \omega_n)
    &= \begin{pmatrix}
        \Sigma_0(\imu\omega_n) \hat 1 & \Delta_0(\imu\omega_n) (\widehat {1\otimes \epsilon})
        \\[1mm]
        \Delta_0(\imu\omega_n) (\widehat {1\otimes \epsilon^{\rm T}}) & \Sigma_0(\imu\omega_n) \hat 1 
    \end{pmatrix},
\end{align}
where the phase of the superconducting order parameter is fixed as real.
We have also introduced the matrix notations such as $\displaystyle (\widehat {1\otimes \epsilon})_{\gm\sg,\gm'\sg'} = \delta_{\gm\gm'} \epsilon_{\sg\sg'}$. 
As for the $O(h^1)$ contribution, we consider Eq.~\eqref{eq:perturb1} and express $\check{G}_1$ on the right-hand side in terms of $\check{\Sigma}_1$ using Eqs.~\eqref{eq:g_ext_gen} and \eqref{eq:g_ext_expand}, thereby formulating a self-consistent equation for the first-order self-energies. As a starting point, we set $\check{\Sigma}_1$ to zero on the right-hand side of this equation. 
The first-order correction to the self-energy then takes the following form:
\begin{align}
    \check {\Sigma}_1^X(\imu \omega_n)
    &= \begin{pmatrix}
        \Sigma^X_1 (\imu\omega_n) \hat X & \Delta^X_1(\imu\omega_n) \hat X (\widehat {1\otimes \epsilon})
        \\[1mm]
        \Delta^X_1(\imu\omega_n) (\widehat {1\otimes \epsilon^{\rm T}})\hat X & -\Sigma^X_1(\imu\omega_n) \hat X^{\rm T} 
    \end{pmatrix} 
    ,
    \label{eq:perturb1_spec}
\end{align}
where $\hat X = \widehat{1\otimes \sigma^z} \equiv \hat{\mathcal S}^z$ for the spin susceptibility and $\hat X = \widehat{\lambda^2 \otimes 1}\equiv \hat{\mathcal L}^z$ for the magnetic orbital susceptibility.
For the derivation, it is useful to note the relations in Appendix~\ref{sec:relation_green_function}.
In the following, we put a label $X$ for the quantities which depend on the choice of $X=\mathcal S^z$ (spin Zeeman) or $X=\mathcal L^z$ (orbital Zeeman).

The form of Eq.~\eqref{eq:perturb1_spec} remains valid if we recover the first-order self-energies in Eq.~\eqref{eq:perturb1}, as confirmed by substituting Eq.~\eqref{eq:perturb1_spec} into the right-hand side of Eq.~\eqref{eq:perturb1}.
Thus, the unknown variables are $\Sigma_{0,1}$ and $\Delta_{0,1}$.
We also note that, in our setup, we have the relations ${\rm Re\,}\Sigma_0(\imu\omega_n)={\rm Im\,}\Delta_0(\imu\omega_n)={\rm Im\,}\Sigma_1(\imu\omega_n)={\rm Re\,}\Delta_1(\imu\omega_n)=0$.

At this point, let us explain the types of Cooper pairs  \cite{Tanaka12,Linder19}.
Without application of the fields, the orbital-symmetric spin-singlet pair is originally formed below the transition temperature.
Under the spin Zeeman field for $X=\mathcal S^z$, the symmetry in spin space is lowered to result in the induction of spin-triplet pairs. Since the spatially local ($s$-wave) pair is considered and the orbital part remains unchanged, i.e. orbital-symmetric, the time-dependence must have an odd functional form due to the Pauli principle, which is called odd-frequency pairing \cite{Tanaka12}.
A similar consideration is applied also to the orbital Zeeman field ($X=\mathcal L^z$), and the corresponding induced odd-frequency pairs are identified as ($s$-wave) orbital-asymmetric spin-singlet Cooper pairs.

A set of equations can now be derived by straightforward calculations. 
For the zeroth-order part, we obtain
\begin{align}
    \Sigma_0(\imu\omega_n) &= - T\sum_{n'} I(\imu\omega_n - \imu\omega_{n'}) G_0 (\imu\omega_{n'}),
    \\
    \Delta_0(\imu\omega_n) &= T\sum_{n'} I(\imu\omega_n - \imu\omega_{n'}) F_0 (\imu\omega_{n'}),
    \\
    G_0(\imu\omega_n) &= \int \diff \ep \rho(\ep) \mathsf g(\ep,\imu\omega_n),
    \\
    F_0(\imu\omega_n) &= \int \diff \ep \rho(\ep) \mathsf f(\ep,\imu\omega_n),
\end{align}
where 
\begin{align}
    I(\imu\nu_m) &= \frac{2}{3}g_0^2 D_{00}(\imu\nu_m) + \frac{10}{3} g_1^2 D_{11}(\imu \nu_m),
    \\
    \mathsf g(\ep,\imu\omega_n) &= \frac{\zeta(\imu\omega_n) + \ep}{\zeta(\imu\omega_n)^2 - \ep^2 - \Delta_0(\imu\omega_n)^2},
    \\
    \mathsf f(\ep,\imu\omega_n) &= \frac{\Delta_0 (\imu\omega_n) }{\zeta(\imu\omega_n)^2 - \ep^2 - \Delta_0(\imu\omega_n)^2},
    \\
    \zeta(\imu\omega_n) &= \imu\omega_n - \Sigma_0(\imu\omega_n).\phantom{\frac{3}{2}}
\end{align}
The phonon Green's function $D_{\eta\eta'}(\imu\nu_m)$ is defined in Eqs.~\eqref{eq:def_phonon_Green_Matsubara} and \eqref{eq:d0}. 
We also obtain the equations for the first-order part as
\begin{align}
    \Sigma_1^X(\imu\omega_n) &= - T\sum_{n'} I_X(\imu\omega_n - \imu\omega_{n'}) G_1^X (\imu\omega_{n'}),
    \label{eq:Sigma1}
    \\
    \Delta_1^X(\imu\omega_n) &= T\sum_{n'} I_X(\imu\omega_n - \imu\omega_{n'}) F^X_1 (\imu\omega_{n'}),
    \label{eq:Delta1}
    \\
    G^X_1(\imu\omega_n) &= \int \diff \ep \rho (\ep) \big[ (-h +\Sigma_1^X) (\mathsf g^2+\mathsf f^2) +2\mathsf g\mathsf f \Delta_1^X \big]
    \label{eq:G1}
    \\
    &\equiv
    \int \diff \ep \rho(\ep)
    \mathsf g_1^X(\ep,\imu\omega_n), 
    \\
    F_1^X(\imu\omega_n) &= \int \diff \ep \rho (\ep) \big[ (-h +\Sigma_1^X)(\mathsf g+\bar {\mathsf g}) \mathsf f  + (\mathsf g\bar {\mathsf g}+\mathsf f^2)\Delta_1^X \big],
    \label{eq:F1}
\end{align}
where
\begin{align}
    I_X(\imu\nu_m) &= \frac{2}{3}g_0^2 D_{00}(\imu\nu_m) + A_X g_1^2 D_{11}(\imu \nu_m),
    \\
    \bar {\mathsf g}(\ep,\imu\omega_n) &= - \mathsf g (\ep,-\imu\omega_n).
\end{align}
The arguments are omitted on the right-hand side of Eqs.~\eqref{eq:G1} and \eqref{eq:F1}.
The constant $A_X$ is defined by 
\begin{align}
    \sum_{\eta=1,3,4,6,8} \hat \lambda^\eta \hat X \hat \lambda^\eta &=  A_X \hat X, 
\end{align}
which gives
$A_{X=\mathcal S^z} = 10/3$ and $A_{X=\mathcal L^z} = - 5/3$.
The latter value originates from the non-commutativity of the asymmetric matrix $\lambda^2$ and symmetric matrix $\lambda^{1,3,4,6,8}$.

In Eqs.~\eqref{eq:G1} and \eqref{eq:F1}, there are contributions proportional directly to the external field $h$, and also the terms with $\Sigma_1^X$ and $\Delta_1^X$ ($\propto h$).
These first-order self-energies originate from the change through the electron-phonon interactions and are called `internal' field corrections.
This is a superconductivity analog of the self-energy corrections to the susceptibility in the Fermi liquid theory \cite{Yamada90,Nakano91}.

Thus, the equations are closed within the framework of linear response theory.
Once the Green's function is obtained, the susceptibility is given by
\begin{align}
    \chi_X = 4T \sum_n G_1^X(\imu\omega_n)/h.
\end{align}
This is based on the $\ep$-integral first approach analogous to Eq.~\eqref{eq:chi_stat_matsu}.
Alternatively, one can also take the Matsubara-sum first approach analogous to Eq.~\eqref{eq:chi_stat_epsil}, which gives 
\begin{align}
    \chi_X &= \frac{4}{h} \int \diff\ep \rho(\ep )G_1^{X\rm (i)}(\ep),
    \\
    G_1^{X\rm (i)}(\ep) &= T\sum_n g_1^X (\ep, \imu\omega_n).
\end{align}

\subsubsection{Numerical results: Temperature dependence \label{sec:numerical_tmp}}

We now numerically solve the set of equations derived in the previous subsection.
Since the phonon self-energy is irrelevant for the qualitative aspects, 
we focus on the phonon Green's function without self-energy, as noted in Sec.~\ref{sec:linear_response}. 
For simplicity, we choose $\omega_0=\omega_1$, and consider the the following effective interaction:
\begin{align}
    I(\imu\nu_m) &= (\lambda_0^{\rm st} + 5\lambda_1^{\rm st})\frac{\omega_0^2}{(\imu\nu_m)^2  - \omega_0^2},
    \label{eq:I_numerics}
    \\
    I_X(\imu\nu_m) &= \qty( \lambda^{\rm st}_0 + \frac{3}{2}A_X \lambda^{\rm st}_1 ) \frac{\omega_0^2}{(\imu\nu_m)^2  - \omega_0^2},
    \label{eq:I_X_numerics}
\end{align}
where $\lambda^{\rm st}_\eta = 4g_\eta^2/3\omega_0$ is a static phonon-mediated interaction for the channel $\eta$.

To elucidate the role of the normal self-energy, we perform numerical calculations for two setups:
\begin{align}
\begin{matrix}
   \text{Setup A:} &\ \   \Sigma_{0},\Sigma_{1}^X \neq 0,\ \ 
   &\Delta_{0}, \Delta_1^X \neq0,
   \\[1mm]
   \text{Setup B:} & \ \  \Sigma_{0}=\Sigma_{1}^X = 0,\ \ 
   &\Delta_{0}, \Delta_1^X\neq0.    
   \label{eq:setup_def}
\end{matrix}
\end{align}
Setup B can be regarded as a simplified version of Setup A, which neglects the effects of the normal self-energy.
This procedure may also be understood in terms of the renormalization of the energies by defining the renormalization factor $Z$ by $\zeta(\imu\omega_n) = Z(\imu\omega_n)^{-1} \imu\omega_n$ \cite{Scalapino_book} and rescaling the physical quantities.

\begin{figure}[tb]
    \centering
    \includegraphics[width=85mm]{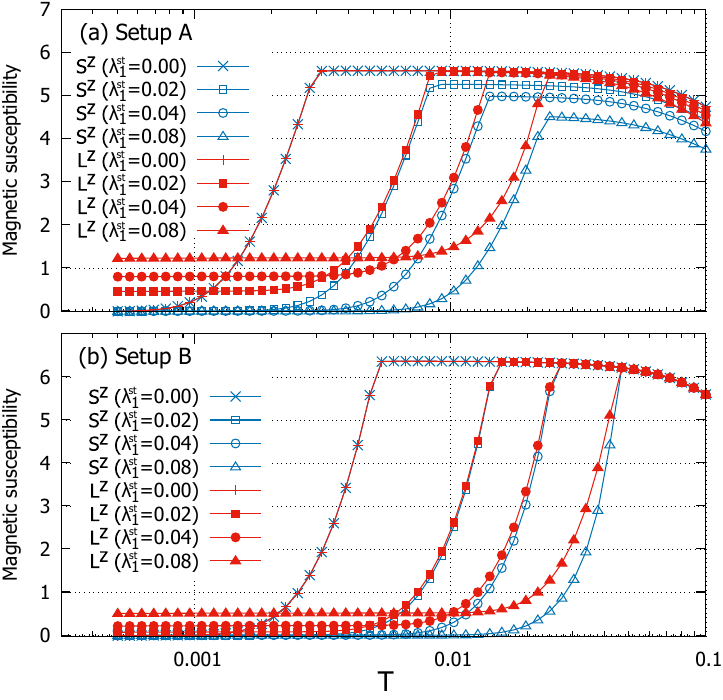}  
    \caption{
Temperature dependence of the magnetic spin ($X=\mathcal S^z$) and orbital ($X=\mathcal L^z$) susceptibilities for (a) the Setup A [with normal self-energy, see Eq.~\eqref{eq:setup_def}] and (b) the Setup B (without normal self-energy). 
    }
    \label{fig:suscep_LRT}
\end{figure}

Figure~\ref{fig:suscep_LRT}(a) shows the temperature dependence of the spin and magnetic orbital susceptibilities for Setup A with $\lambda_0^{\rm st} = 0.2$ eV.
The behaviors are basically similar to those in Fig.~\ref{fig:Sz-Lz_Tdep}:
The magnetic orbital susceptibility ($X=\mathcal L^z$) remains nonzero in the presence of the multiorbital effect ($\lambda_1^{\rm st}\neq 0$) at low temperatures, while the spin susceptibility ($X=\mathcal S^z$) becomes zero.
As for the Setup B shown in Fig.~\ref{fig:suscep_LRT}(b), although the results are quantitatively different due to the absence of the normal self-energy, the qualitative behaviors remain unchanged.
Hence, the normal self-energy has only a secondary effect, and the anomalous self-energy plays an essential role for the remaining magnetic orbital susceptibility and vanishing spin susceptibility at low $T$.

Note that, at $\lambda^{\rm st}_1 = 2\lambda^{\rm st}_0 / 5 = 0.08$ eV, 
the first-order contribution vanishes due to $I_{X=\mathcal L^z} (\imu\nu_m) = 0$ for the magnetic orbital response.
In this particular case, there is no internal field contribution and the discussion at the end of Sec.~\ref{sec:review_bcs} can be applied.
Then, 
 it is evident that the magnetic orbital susceptibility becomes nonzero because of the retardation effect, which is caused by the absence of the high-frequency ($ \gtrsim \omega_{0}$) contribution from the anomalous part. 
 On the other hand, for the spin susceptibility, $I_{X=\mathcal S^z}$ remains and is identical to $I$, i.e., the same interaction kernel is shared between the zeroth- and first-order equations. However, the reason for the absence of the spin susceptibility is nontrivial, and the following discussion is mainly devoted to its clarification.
 Since orbital degrees of freedom are not involved directly in the spin susceptibility, the conclusion should be applied to a wide range of $s$-wave spin-singlet superconductors with a retardation effect.

\begin{table*}[t]
\begin{tabular}{|c||c|c|c|}
\hline
\hspace{2mm}
{\bf Contributions to $\bm \chi_{\bm X}$ }
\hspace{2mm}
&\hspace{2mm} (a) BCS [No retardation]\hspace{2mm} & \hspace{2mm}(b) Eliashberg ($X=\mathcal S^z$) \hspace{2mm} & \hspace{0mm} (c) Eliashberg ($X=\mathcal L^z$)\hspace{0mm}   \\[2pt]
\hline
\hline 
Positive contribution
&  \multicolumn{3}{c|}{Normal part (Pauli paramagnetism)}\\[2pt] \cline{1-4}
\multirow{2}{*}{\shortstack{\\ Negative contributions \\ from pair potentials}}  & \multirow{2}{*}{Spin-singlet pair} & Spin-singlet pair &  Spin-singlet pair   \\[2pt]
 &   & $+$ Odd-freq. pair &  $+$ Odd-freq. pair (diminished) 
\\[2pt] \cline{1-4}
Total &  zero & zero &  nonzero\\[2pt] \cline{1-4}
\hline
\end{tabular}
\caption{
Contributions to the magnetic susceptibility for (a) the spin-Zeeman effect in BCS theory (no retardation), (b) the spin Zeeman effect in Eliashberg theory, and (c) the orbital Zeeman effect in multiorbital Eliashberg theory. 
The primary Cooper pair has the structure of an (even-frequency) orbital-symmetric spin-singlet state, whereas the secondarily induced odd-frequency Cooper pair, generated by the magnetic field, is an orbital-symmetric spin-triplet state for (b) $X=\mathcal S^z$, and an orbital-asymmetric spin-singlet state for (c) $X=\mathcal L^z$.
The contribution from the odd-frequency pair potential in (c) is diminished compared to the case (b) by a multiorbital effect. 
\label{tab:contribution}
}
\end{table*}

\subsubsection{Numerical results: Frequency dependence 
\label{sec:numerical_freq}}

\begin{figure}[tb]
    \includegraphics[width=85mm]{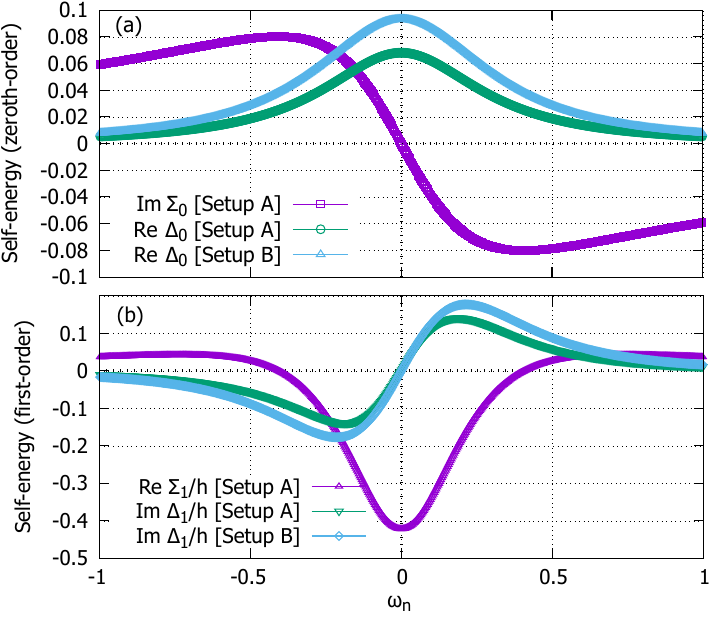}  
    \caption{
Normal and anomalous self-energies with the choice $X=\mathcal S^z$ at $T=0.001 $ eV  
for (a) zeroth-order and (b) first-order contributions with respect to the external field $h$.
Note that 
the not plotted quantities are zero: ${\rm Re\,}\Sigma_0(\imu\omega_n)={\rm Im\,}\Delta_0(\imu\omega_n) = 0$ in (a) and ${\rm Im\,}\Sigma_1^X(\imu\omega_n) = {\rm Re\,}\Delta_1^X(\imu\omega_n) = 0$ in (b). 
    }
    \label{fig:Self-energies_LRT}
\end{figure}

Figure~\ref{fig:Self-energies_LRT} shows the Matsubara frequency dependence of the normal and anomalous self-energies at $T=0.001$ eV and $\lambda_1^{\rm st}=0.08$ eV for $X=\mathcal S^z$ (spin Zeeman).
In Fig.~\ref{fig:Self-energies_LRT}(a), the zeroth-order contributions are displayed.
The behavior is a standard one: The anomalous self-energy decays for $|\omega_n|\gg \omega_0$, since the electron-phonon interaction becomes ineffective at high energy.
We see that the anomalous parts are similar between the Setup A and B, demonstrating the irrelevance of the normal self-energy for qualitative aspects.

Figure~\ref{fig:Self-energies_LRT}(b) shows the $O(h^1)$ corrections to the normal and anomalous self-energies for $X=\mathcal S^z$.
The normal part has an even-function shape and is interpreted as a correction to the external field $h$, which is modified to $\tilde h = h - {\rm Re\,}\Sigma_1$ [see the right hand sides of Eqs.~\eqref{eq:G1} and \eqref{eq:F1}].
As for the anomalous part, the functional form is odd with respect to the Matsubara frequency.
Hence, this can be regarded as an odd-frequency ($s$-wave, spin-triplet) pair induced by the external magnetic field \cite{Bergeret05,Tanaka12, 
Matsumoto12, Fukui18}. 
Here again, the anomalous part is similar between the Setups A and B, and hence the normal self-energy may be neglected when studying the qualitative behaviors.
Since the odd-frequency pair potential $\Delta_1^X$ contributes to the susceptibility, as is evident from Eq.~\eqref{eq:G1}, the presence of the odd-frequency (spin-triplet) pairs is essential for the vanishing spin susceptibility: without these pairs, the spin susceptibility does not become zero for spin-singlet superconductors. 
This situation is summarized in Tab.~\ref{tab:contribution}~(b), which is compared to the standard BCS case without retardation as listed in Tab.~\ref{tab:contribution}~(a).
We note that the frequency-dependent pair potential is not induced for the non-retarded electronic interaction $I_X(\imu\nu_m) = {\rm const.}$ according to Eq.~\eqref{eq:Delta1}.

As for the orbital Zeeman effect ($X=\mathcal L^z$), on the other hand, the induction of the odd-frequency pair (orbital-asymmetric/spin-singlet) is smaller than the spin Zeeman effect due to the relation $- I_{X=\mathcal S^z} > -I_{X=\mathcal L^z}$ [see Eq.~\eqref{eq:I_X_numerics} with $A_{X=\mathcal S^z} = 10/3$ and $A_{X=\mathcal L^z} = - 5/3$].
Hence, the contribution from odd-frequency pair is smaller compared to the spin-Zeeman case, which results in the nonvanishing susceptibility for the orbital Zeeman case [see Tab.~\ref{tab:contribution}~(c)].
In our simplified model, the magnitude of the odd-frequency pairs can even become zero for $\lambda_1^{\rm st} = \frac{2}{5} \lambda_0^{\rm st}$. 
If we further consider the case with $\lambda_1^{\rm st} > \frac{2}{5} \lambda_0^{\rm st}$, the odd-frequency pairs positively contribute to the magnetic susceptibility (not shown in Tab.~\ref{tab:contribution}).

In this way, the odd-frequency pairing plays an important role for the magnetic susceptibilities, which is one of the main conclusions of this paper.
We emphasize that the mechanism of vanishing spin susceptibility in the presence of retardation is applied to single-orbital superconductors and is not specific to multiorbital systems. In addition, the remaining orbital susceptibility is expected for a generic multiorbital superconductor coupled to anisotropic vibrational modes. Hence, our conclusion should be applied to a wide range of superconductors with retardation effects.

In the remainder of this subsection, 
we further discuss the spectral decomposition of the magnetic susceptibility and explore what kind of cancellations lead to a vanishing susceptibility. 
For this purpose, we focus on 
the Setup B [see Eq.~\eqref{eq:setup_def}].
First of all, we show the expression of the susceptibility for the Setup B by performing the $\ep$-integral:
\begin{widetext}
    \begin{align}
    \chi_X &= 4\frac{T}{h} \sum_n \qty[ \Lambda'(\imu\tilde \Omega_n) h + \qty( - \Lambda'(\imu\tilde \Omega_n) + \frac{\Lambda(\imu\tilde \Omega_n)}{\imu\tilde \Omega_n} ) 
    \frac{\Delta_0(\imu\omega_n)^2}{\tilde \Omega_n^2}
    \qty( h - \frac{\imu\omega_n \Delta_1^X(\imu\omega_n)}{\Delta_0(\imu\omega_n)} )
    ]
\ \equiv\ \frac{T}{h} \sum_n \qty[ G_1^{\rm ext}(\imu\omega_n) + G_1^{X,{\rm int}}(\imu\omega_n)], 
\end{align}
\end{widetext}
where we have defined $    \tilde \Omega_n=  \sqrt{\omega_n^2 + \Delta_0(\imu\omega_n)^2}
$.
We have introduced the external-field ($G_1^{\rm ext}$) and internal-field ($G_1^{X,\rm int}$) contributions to the susceptibility, each of which is proportional to $h$ and $\Delta_1$, respectively.
Note that the external-field contribution does not depend on the choice of $X$.

Since we are interested in the effect of retardation through the electron-phonon coupling, let us subtract the static component.
We define the static part of the anomalous self-energy by $\Delta_{\rm st} = \Delta(\imu\omega_n =0)$, whose contribution to the susceptibility is given by Eq.~\eqref{eq:chi_stat_matsu}. 
At low temperature, we rewrite the susceptibility as
\begin{align}
    \chi_X &\simeq \chi_X - \chi_{\rm BCS}(\Delta_{\rm st})
    \\
    &= \frac{T}{h} \sum_n \qty[ \delta G_1^{\rm ext}(\imu\omega_n)  + G_1^{X,\rm int}(\imu\omega_n)], 
    \label{eq:LRT_setup_B_matsu}
\end{align}
where 
we have introduced the retarded contribution for the external-field part by $\delta G_1^{\rm ext}(\imu\omega_n) \equiv G_1^{\rm ext}(\imu\omega_n) - G_1^{\rm (ii)} (\imu\omega_n; \Delta_{\rm st})$
with $G_1^{\rm (ii)}$ defined in Eq.~\eqref{eq:chi_stat_matsu}.
Note that the spin susceptibility in the BCS theory, $\chi_{\rm BCS}(\Delta_{\rm st})$, goes to zero as $\sim (\Delta_{\rm st}/T)^{3/2} \epn^{-\Delta_{\rm st}/T} $ at low $T$.

\begin{figure}[tb]
    \centering
\includegraphics[width=85mm]{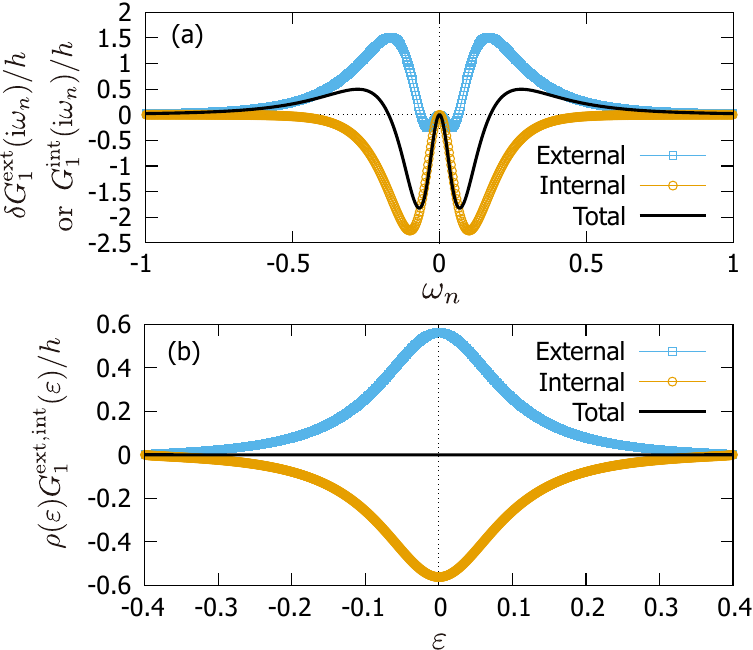}  
    \caption{
External-field induced Green's function defiend by Eq.~\eqref{eq:LRT_setup_B_matsu} for $X=\mathcal S^z$.
We used the $\Delta$-only model for the analysis (i.e., the Setup B).
    }
    \label{fig:decomp_LRT}
\end{figure}

Figure~\ref{fig:decomp_LRT}(a) shows the external and internal contributions as a function of $\omega_n$ defined in Eq.~\eqref{eq:LRT_setup_B_matsu} for $X=\mathcal S^z$.
For the external-field part, the dominant contribution is positive and enters at relatively high frequencies, while the internal contribution is negative and dominantly contributed at low frequencies.
These two contributions are added to give zero at low $T$.

In addition, we can also consider the $\ep$-dependence by taking the Matsubara sum first, as in Eq.~\eqref{eq:chi_stat_epsil}.
Namely, we decompose the susceptibility as
\begin{align}
    \chi_X = \frac{4}{h} \int \diff\ep \rho(\ep ) \Big[ G_1^{\rm (i),ext}(\ep) + G_1^{X\rm (i),int}(\ep) \Big]. 
    \label{eq:LRT_setup_B_epsil}
\end{align} 
Figure~\ref{fig:decomp_LRT}(b) shows the $\ep$-dependence of these quantities.
It is notable that the cancellation of external- and internal-field contributions occurs at each $\ep$.
These behaviors are re-examined based on the analytic calculations in the next subsection.

\subsubsection{Analysis in the strong retardation limit \label{sec:strong_ret}}

With the numerical results provided, it is still not fully clear why the spin ($X=\mathcal S^z$) susceptibility vanishes. 
To gain more insight, let us further examine the mathematical structure of the Eliashberg equation.

Here, by providing the analytic solution, we support the conclusion of the previous subsection that the odd-frequency pairing plays a crucial role in the vanishing of the spin susceptibility.
We work with the Setup B, and take the strong retardation limit $\omega_0 \to 0$.
Although this is not a physically realistic assumption, it makes the mathematical structure simple and allows us to interpret the numerical results in more depth.
In this limit, the interaction in the Matsubara frequency domain is given by
\begin{align}
I(\imu\nu_m) &= - 2\pi C \delta(\nu_m),
\\
I_X(\imu\nu_m) &= - 2\pi C_X \delta(\nu_m),
\end{align}
where $C = \tfrac 2 3 g_0^2 + \tfrac{10}{3} g_1^2$ and $C_X = \tfrac 2 3 g_0^2 + A_X g_1^2$.
Then, for $T\to 0$, the self-energies become
\begin{align}
    \Delta_0(\imu\omega_n) &= -C F_0(\imu\omega_n),
    \\
    \Delta_1^X(\imu\omega_n) &= - C_X F_1(\imu\omega_n).
\end{align}
Using Eqs.~\eqref{eq:G1} and \eqref{eq:F1}, we obtain 
\begin{align}
    \Delta_0(\imu\omega_n) &= \pi\rho_0 C \sqrt{1 -w^2},
    \label{eq:Delta0_analy}
    \\[1mm]
    \Delta_1^X (\imu\omega_n)/h &= 
    \frac{\ \imu w \sqrt{1-w^2}\ }{\displaystyle \frac{C}{C_X}- w^2},
    \label{eq:Delta1_analy}
\end{align}
where $\rho_0 = \rho(0) = 2/\pi D$, 
$\Delta_{\rm st} = \Delta_0(0)=  \pi \rho_0 C$, and $w = \omega_n / \Delta_{\rm st}$.
Note that the anomalous self-energies are zero for $|\omega_n| > \pi \rho_0 C$. 
In this way, it is evident in the $\omega_0 \to 0$ limit that $\Delta_0$ is an even function and $\Delta_1^X$ an odd function in frequency.

With the above results, we can explicitly derive the magnetic susceptibilities as
\begin{align}
   \chi_X &= 4\rho_0 \  \mathcal I \qty( \frac{C}{C_X})
   ,
\end{align}
where we have introduced
\begin{align}
    \mathcal I(x) &= 
    1 - \int_0^1 \diff w \frac{1-w^2}{x-w^2}
=
    \frac{x-1}{2\sqrt x} \ln \frac{\sqrt x + 1}{\sqrt x - 1}
    .
\end{align}
This function behaves as $\mathcal I \to 0$ for $x\to 1$ and $\mathcal I \to 1$ for $x\to \infty$.
The dimensionless constant is given by
\begin{align}
    \frac{C}{C_{X=\mathcal L^z}}
    &= \frac{1}{\displaystyle \ 1- \frac{15 \lambda_1^{\rm st}}{\  2 \lambda_0^{\rm st} + 10 \lambda_1^{\rm st}\  }\ }
    .
\end{align}
For $X=\mathcal S^z$, we have $C/C_X = 1$.

Thus, we have succeeded to account for the fact that the orbital magnetic susceptibility ($X=\mathcal L_z$) is nonzero in the presence of the coupling to the Jahn-Teller phonon.
For $\lambda_1^{\rm st} \ll \lambda_0^{\rm st}$, we can explicitly evaluate the asymptotic form as
\begin{align}
    \chi_{X=\mathcal L^z} &\simeq - \frac{15\rho_0 \lambda_1^{\rm st}}{\lambda_0^{\rm st}}
    \ln \frac{15\lambda_1^{\rm st}}{8\lambda_0^{\rm st}}.
\end{align}
On the other hand, for the spin susceptibility ($X=\mathcal S^z$), we obtain $\chi_{X=\mathcal S^z}=0$ from the condition $C_{X=\mathcal S^z} = C$.
In other words, the spin susceptibility vanishes if the same interaction kernel is shared between the zeroth-order [$I(\imu \nu_m)$] and first-order [$I_X(\imu \nu_m)$] equations.

\begin{figure}[tb]
    \centering
\includegraphics[width=80mm]{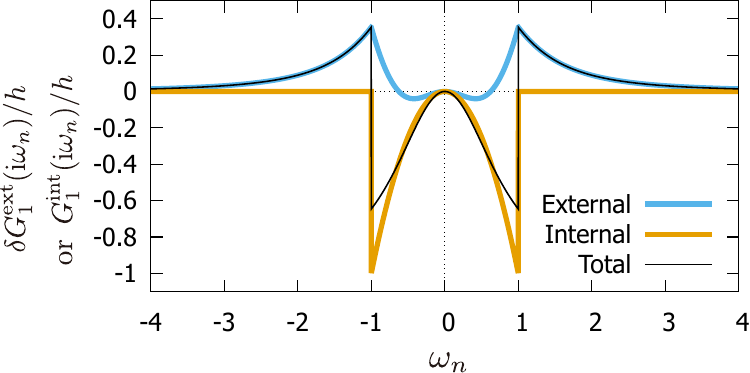}  
    \caption{
    Analytically obtained
external-field induced Green's function defined by Eqs.~\eqref{eq:G1ext_analy} and \eqref{eq:G1int_analy} for $X=\mathcal S^z$.
The vertical axis values are normalized by $\pi \rho_0 h/\Delta_{\rm st}$, and the horizontal ones by $\Delta_{\rm st}$.
We have used the Setup B (no normal self-energy) for the analysis.
    }
    \label{fig:decomp_LRT_analy}
\end{figure}

The spectral decomposition for the susceptibility can also be analytically performed, and Eqs.~\eqref{eq:LRT_setup_B_matsu} and \eqref{eq:LRT_setup_B_epsil} can be explicitly evaluated.
We obtain
\begin{align}
    \delta G_1^{\rm ext} (\imu\omega_n)
    &= \frac{\pi \rho_0 h}{\Delta_{\rm st}} \left\{
\begin{matrix}
(1+w^2)^{-3/2} - (1-w^2) & \  & (|w|<1)
\\[1mm]
(1+w^2)^{-3/2} & \  & (|w|>1)
\end{matrix}
    \right.,
    \label{eq:G1ext_analy}
\\
G_1^{X,\rm int} (\imu\omega_n) &= - \frac{\pi \rho_0 h}{\Delta_{\rm st}} \ \frac{ \  w^2(1-w^2) \ }{\displaystyle \frac{C}{C_X} - w^2},
\label{eq:G1int_analy}
\end{align}
which are plotted in Fig.~\ref{fig:decomp_LRT_analy} for $X=\mathcal S^z$.
This qualitatively reproduces the functional form of Fig.~\ref{fig:decomp_LRT}(a).
We also perform another spectral decomposition as
\begin{align}
    G_1^{\rm (i), ext} (\ep)
    &= - G_1^{X=\mathcal S^z,\rm (i), int} (\ep)
    \\
    &= \frac{2h}{3\pi \Delta_{\rm st}}
    \ \frac{1}{1+(\ep/\Delta_{\rm st})^2}
    ,
\end{align}
which is consistent with Fig.~\ref{fig:decomp_LRT}(b).

\subsubsection{
Alternative interpretation of odd-frequency pairing
}
\label{sec:TimeIndep}

In the preceding sections, we have demonstrated that odd-frequency pairing makes a significant contribution to the magnetic susceptibilities. However, since the odd-frequency pair amplitude is inherently a dynamical quantity, it is not straightforward to develop an intuitive understanding of it. 
To gain further insight, we now turn to an alternative representation. 
In principle, dynamical quantities can be represented by static quantities using equations of motion, which helps understanding the origin of dynamical behaviors \cite{Emery92,Balatsky93}.
To explicitly see this property for odd-frequency pairing, we begin with the local ($s$-wave) spin-triplet pair amplitude induced by the spin Zeeman field ($X=\mathcal S^z$):
\begin{align}
    F_{\rm trip} (\tau) &=  - \sum_{\gm\sg\sg'} \la \mathcal T c_{i\gm \sg}^\dg(\tau) (\sg^z \epsilon)_{\sg\sg'} c^\dg_{i\gm\sg'} \ra
    .
    \label{eq:odd-freq_def}
\end{align}
Note that this imaginary-time dependent pair {\it amplitude} corresponds to the pair {\it potential} due to the self-consistent relation in Eq.~\eqref{eq:Delta1}.
This correspondence is valid if the retarded interaction kernel $I_X$ is nonzero.
We note that it is not applicable to the BCS theory, 
where an odd-frequency pair amplitude (= anomalous Green's function) is induced but the odd-frequency pair potential (= anomalous self-energy) is zero. 
According to the discussion in the previous subsections, the quantity that contributes to the magnetic susceptibility is the odd-frequency potential $\Delta_1$ (see Tab.~\ref{tab:contribution}).
We emphasize that the pair amplitudes in the absence of the pair potential can also contribute physical quantities such as transport coefficients in junctions \cite{Bergeret05,Tanaka12}.
With these considerations in mind, we proceed to examine the properties of the odd-frequency pair amplitudes.

We may represent Eq.~\eqref{eq:odd-freq_def} as a Taylor expansion:
\begin{align}
    &F_{\rm trip} (\tau>0) = \sum_{k=0}^\infty A_{\rm trip}^{(k)} \tau^k
    ,
    \\
   &A_{\rm trip}^{(k)} = (-1)^{k+1}\sum_{\gm\sg\sg'} 
    (\sg^z \epsilon)_{\sg\sg'} \big\la (\mathscr L^k c_{i\gm \sg}^\dg) c^\dg_{i\gm\sg'} \big\ra
    ,
\end{align}
where we have defined the Liouvillian $\mathscr L O = [O,\mathscr{H}]$.
The first-order coefficient, which gives the dominant contribution at small (imaginary-)time, 
is given by
\begin{align}
    A_{\rm trip}^{(1)}
    &= 6h F_{\rm sing}
    ,
\end{align}
where we have introduced the static spin-singlet pair amplitude $F_{\rm sing}$ by $\la c^\dg_{i\gm\sg} c^\dg_{i\gm'\sg'} \ra = F_{\rm sing}\delta_{\gm\gm'} \epsilon_{\sg\sg'}$.
Note that the commutator with the hopping term and electron-phonon coupling term vanishes due to the inversion symmetry and the Pauli principle (the Coulomb interaction is neglected in this subsection) 
\footnote{
The commutation relation with interaction term in general produces the composite pair amplitude to characterize the short-time behavior of the odd-frequency pairing \cite{Emery92,Balatsky93,Hoshino14_2}
}.
In this way, the odd-frequency triplet pairing is characterized by the combination of the Zeeman field $h$ and the spin-singlet pair amplitude $F_{\rm sing}$.

In the Matsubara frequency domain, we utilize the high-frequency expansion in the form
\begin{align}
    &F_{\rm trip} (\imu\omega_n) = \sum_{k=1}^\infty \frac{B_{\rm trip}^{(k)}}{(\imu\omega_n)^{k}}
    ,
    \\
    &B_{\rm trip}^{(k)} = \sum_{\gm\sg\sg'} 
    (\sg^z \epsilon)_{\sg\sg'} \Big\la \big\{ ( \mathscr L^{k-1} c_{i\gm \sg}^\dg), c^\dg_{i\gm\sg'}\big\} \Big\ra
    ,
\end{align}
where $\{a,b\}=ab+ba$ is the anticommutator.
In the electron-phonon coupled system without Coulomb interaction 
\footnote{
The third-order component exists in the presence of the Coulomb repulsive interaction, since a static pair potential is secondarily induced \cite{Kaga22}
},
the leading-order contribution is given by the fifth-order component:
\begin{align}
    B^{(5)}_{\rm trip}
    &= 32hF_{\rm sing} \sum_\eta \omega_\eta g_\eta^2
    .
\end{align}
Here again, both the Zeeman field and spin-singlet pair amplitude are involved in the coefficient relevant to odd-frequency pairing.

A similar discussion can be provided also for the odd-frequency pairing under the orbital magnetic field ($X=\mathcal L^z$), where orbital asymmetric spin-singlet pairs are induced by the external field.
As for the high-frequency expansion, a more complicated expression is obtained due to the non-commutative nature of the Gell-Mann matrices.

\section{Summary \label{sec:summary}}

Alkali-metal-doped fullerides are multiorbital, strongly correlated electron systems with Jahn-Teller phonons that produce an effectively antiferromagnetic (AFM) Hund’s coupling. While this effective static coupling simplifies theoretical treatments, the actual electron–phonon interaction exhibits retardation effects that must be assessed to accurately describe superconducting and orbital-ordered states.

In this work, we investigated these retardation effects using dynamical mean-field theory combined with multiorbital Eliashberg theory, appropriate for the weak-coupling regime where superconductivity is observed. Unlike previous studies focusing only on superconductivity, we extended the Eliashberg framework to calculate spin, orbital, and superconducting susceptibilities. Our analysis shows that, while the orbital electric quadrupole and $s$-wave superconducting susceptibilities behave similarly under static (i.e., effective AFM Hund's coupling) and retarded interactions, significant qualitative differences can arise in the orbital magnetic susceptibility.

We further examined the magnetic susceptibilities in the superconducting state with $s$-wave spin-singlet pairing. Despite the spin-singlet nature, the magnetic susceptibility remains nonzero due to multiorbital and retardation effects. We showed that this residual response is deeply connected to the induction of an odd-frequency pairing by the magnetic field. 
In the strong-retardation limit, we have also obtained an analytic solution in the low-temperature limit.
Our results 
provide new insight into the microscopic origin of the magnetic susceptibilities for superconductors with retardation effects.

\section*{Acknowledgement}

We would like to thank Hiroaki Ikeda for useful discussions.
N.O. was supported by Takafumi Horikawa Educational Foundation and Futaba Foundation. 
This work was supported by JSPS KAKENHI Grants 
No.~JP24K00578,
No.~JP24K00583,
No.~JP23K25827,
No.~JP23K17668, 
No.~JP23H04869, 
No.~JP23H04519, 
No.~JP23K03307,
``Program for Promoting Researches on the Supercomputer Fugaku'' (Project ID: JPMXP1020230411) from MEXT, 
and by
the Grant-in-Aid for Transformative Research Areas (A) 
“Correlation Design Science” (KAKENHI Grant 
No.~JP25H01249)
from JSPS of Japan.\\

\appendix

\section{Calculation method for the local electron Green's function \label{sec:calc_green_function_app}}

For the local electronic Green's function in the $12\times 12$ Nambu matrix form, we need to evaluate 
the following integral 
\begin{align}
    \check G(z) = \int \diff \ep \rho(\ep) [z \check 1 -(\ep - \mu) \check \tau_3 - \check \Sigma(z)]^{-1}, 
\end{align}
where $z$ ($=\imu \omega_n$) is the complex frequency.
We want to utilize the Hilbert transformation formula
\begin{align}
    \Lambda(z) &= \int \frac{\rho(\ep)}{z-\ep} \diff \ep, 
\end{align}
which can be evaluated analytically for a semi-elliptical $\rho(\ep)$ [cf. Eq.~\eqref{eq:dos_analytical}].
Since $\check \tau_3$ and $\check \Sigma (z)$ are not diagonalizable simultaneously, we rewrite 
the expression by introducing the auxiliary quantity
\begin{align}
    \check \zeta (z) = z\check 1 + \mu \check \tau_3 - \check \Sigma(z).
\end{align}
The Green's function matrix is then given by
\begin{align}
    \check G(z) = \int \diff \ep \rho(\ep) [\check \tau_3\check \zeta(z)  -\ep \check 1 ]^{-1} \check \tau_3.
\end{align}
We diagonalize the matrix as
\begin{align}
    \check \tau_3\check \zeta(z)
    = \check V(z) \check L(z) \check V^{-1}(z). 
\end{align}
The elements of the Green's functions are thus evaluated as
\begin{align}
    \check G_{ij}(z) &= \sum_\al \check V_{i\al} \  g\big( \check L_{\al\al}(z) \big) 
    \ [\check V^{-1}\check \tau_3]_{\al j},
\end{align}
where $\al$ is a diagonalized basis.

\section{Relations among the Green's functions \label{sec:relation_green_function} 
}
It is useful to summarize the general relations among the Green's functions.
In the matrix representation, we have the relations
\begin{align}
\check G(\tau)^* &=
\check G(\tau)^{\rm T}, 
\\
\check G(- \tau) &= -
\begin{pmatrix}
\hat {\bar G}(\tau)^{\rm T} & \hat F(\tau)^{\rm T} \\
    \hat {\bar F}(\tau)^{\rm T} & \hat G(\tau)^{\rm T}    
\end{pmatrix}
= - \check \tau_1 \check G(\tau)^{\rm T} \check \tau_1, 
\label{eq:minus_tau_to_plus_tau}
\end{align}
where
\begin{align}
    \check \tau_1 &= 
    \begin{pmatrix}
        \hat 0 & \hat 1 \\
        \hat 1 & \hat 0
    \end{pmatrix}
\end{align}
is the Pauli matrix in the Nambu space.
In the Matsubara frequency domain, we have 
\begin{align}
    \check G (\imu\omega_n)^* 
    &= \check G(-\imu\omega_n)^{\rm T}, 
    \nonumber \\
\check G(- \imu \omega_n) &= -
\begin{pmatrix}
\hat {\bar G}(\imu \omega_n)^{\rm T} & \hat F(\imu \omega_n)^{\rm T} \\
    \hat {\bar F}(\imu \omega_n)^{\rm T} & \hat G(\imu \omega_n)^{\rm T}    
\end{pmatrix}
    = - \check \tau_1 \check G(\imu\omega_n)^{\rm T} \check \tau_1. 
\end{align}
A set of similar relations also exists for the self-energies.

\bibliography{perturbation}
\bibliographystyle{apsrev4-2}

\end{document}